\documentstyle[preprint,eqsecnum,aps,pra]{revtex}

\begin{document}
\renewcommand{\thefootnote}{\fnsymbol{footnote}}
\draft
\title{Influence of through-flow on linear pattern formation properties in
binary mixture convection}
\author{Ch.~Jung, M.~L\"{u}cke, and P.~B\"{u}chel \\}
\address{Institut f\"{u}r Theoretische Physik,Universit\"{a}t des Saarlandes,
 D-66041 Saarbr\"{u}cken, Germany \\}

\renewcommand{\thefootnote}{\arabic{footnote}}
\setcounter{footnote}{0}
\date{\today}
\maketitle

% ABSTRACT

\begin{abstract}

We investigate how a horizontal plane Poiseuille
shear flow changes linear convection properties
in binary fluid layers heated from below. The full linear
field equations are solved with a shooting method 
for realistic top and bottom boundary conditions. Relevant
characteristic growth exponents and the spatial structure
of their associated eigenfunctions are evaluated for different 
perturbations of the conductive state. Through-flow 
induced changes of the bifurcation thresholds (stability
boundaries) for different types of convective solutions
are determined in the control parameter space spanned
by Rayleigh number, Soret coupling (positive as well as negative),
and through-flow Reynolds number. We elucidate the 
through-flow induced lifting of the Hopf symmetry degeneracy
of left and right traveling waves in mixtures with negative 
Soret coupling.
Finally we determine with a saddle point analysis of the 
complex dispersion relation of the field equations over the
complex wave number plane the borders between absolute and 
convective instabilities for different types of perturbations 
in comparison with the appropriate Ginzburg-Landau amplitude
equation approximation.
 
\end{abstract}


%\begin {multicols}{2}
\narrowtext

% BEGIN OF THE TEXT

%%%%%%%%%%%%%%%%%%%%%%%%%%%%%%%%%%%%%%%%%%%%%%%%%%%%%%%%%%%%%%%%%%%%
% \input{i1004cj.tex}
%%%%%%%%%%%%%%%%%%%%%%%%%%%%%%%%%%%%%%%%%%%%%%%%%%%%%%%%%%%%%%%%%%%%

\section{INTRODUCTION}\label{i} 
An externally imposed flow can influence the spatiotemporal behavior
of dissipative structures growing in forced nonequilibrium systems.
Examples are chemical and reaction-diffusion systems, biological 
problems, and the large variety of different 
hydrodynamic instabilities leading to pattern formation \cite{crho93}.
Here we theoretically investigate with a linear analysis of the relevant 
field equations the spatiotemporal properties of convection solutions
that bifurcate out of the homogeneous, quiescent conductive state in 
a binary fluid layer heated from below.

A lot of experimental \cite{DAC95,KKS94,k93,PES96,GPC85,lp89,LB92,ms91,ZMD92},
analytical \cite{BPS89,BLN86,ClK91,C88,LL87a,linz88,ML88}, 
and numerical \cite{BLKS95,Be91b,DTM87,Y89} activities have been devoted 
recently to investigate these primary convection patterns in the
absence of through-flow \cite{OV}.  They revealed a variety of bifurcation
properties and spatiotemporal behavior that is much richer than
that of stationary mirror symmetric roll patterns growing in the
supercritical bifurcation of the standard Rayleigh B\'enard setup
with a pure fluid, say water. In binary mixtures like, e.g., 
ethanol-water there occur stationary square and roll patterns but
also symmetry degenerate left or right traveling convection waves.
The bifurcation of the traveling wave (TW) solutions
and of the stationary roll patterns can be super-, tri-, or subcritical
relative to the critical heating rate. These different solution properties
are controlled by the combination of thermal forcing --- i.e. the
Rayleigh number --- and the strength $\psi$ \cite{crho93} of the Soret coupling
between temperature and concentration field. It is the concentration field
that causes the rich structure formation behavior via its contribution
to the buoyancy force field which drives convection.

Without lateral through-flow several experimental and
theoretical papers have addressed the linear convection properties
\cite{CL75,GCR79,LLT83,GB86,kp86,wh88,km88,ck88-1,sz92,st91,LLM91,h92,sh95}.
The most comprehensive and most accurate results were obtained
in the more recent numerical work \cite{km88,ck88-1,sz92,st91,LLM91,h92,sh95}.    
In this work we determine how a lateral through-flow changes
the structure and dynamics of convection fields at onset. We 
evaluate relevant characteristic exponents and the spatial structure
 of their associated eigenfunctions for different perturbations of 
the conductive basic state and determine the through-flow 
induced changes of the bifurcation thresholds for different types 
of convective solutions. These thresholds are defined by vanishing 
real parts of the characteristic exponents. They mark stability boundaries
of the conductive state against different convective 
perturbations in the control parameter space spanned by the Rayleigh 
number $Ra$, the Soret coupling $\psi$, and the through-flow Reynolds
number $Re$.
We have numerically solved the full linear field equations subject 
to realistic boundary conditions using a shooting method 
for perturbations with wavevectors parallel to the through-flow. The 
solutions for other wavevectors can be obtained from the former by 
a straightforward symmetry transformation. Results were obtained 
for positive as well as for negative  $\psi$. For $\psi < 0$ the most
 important result is that the through-flow lifts the Hopf symmetry 
degeneracy of left and right traveling waves at $Re = 0$: frequencies, 
bifurcation thresholds, and structural properties of the two waves
 are changed dramatically. Mixtures with more negative  $\psi$ require 
a larger $Re$ for the changes to reach a comparable relative size. 
For sufficiently large $Re$ the lowest relevant bifurcation threshold 
of binary mixtures with any $\psi$ asymptotically approaches the 
critical Rayleigh number $Ra_c(Re, \psi = 0)$ of a pure fluid with 
imposed through-flow.
Then the externally imposed shear flow 
eliminates the Soret-induced coupling effects between the convective 
concentration field and the other fields by suppressing vertical 
convective transport of Soret driven concentration perturbations. 

Our paper is organized as follows.
In Sec.~\ref{ii} we describe the system.
In Sec.~\ref{iii} we review the linearized equations for 
perturbations of the conductive state, their boundary 
conditions, the eigenvalue problem for the characteristic 
exponents, relevant symmetry properties, and the behavior 
for small through-flow rates. 
Sec.~\ref{iv} contains our results concerning bifurcation
 properties --- stability thresholds, wave numbers, frequencies, 
and eigenfunctions --- for negative and positive Soret 
coupling $\psi$ as functions of the through-flow Reynolds 
number $Re$.
In Sec.~\ref{v} we compare borderlines between absolute and 
convective instabilities obtained for different types of 
perturbations from the field equations with results from the
Ginzburg-Landau amplitude equation. The last section gives 
a brief summary of our work.
Appendix~\ref{appa} contains details of our shooting method. 
There we also describe our procedure to find 
saddle points of the dispersion relation of the field equations in 
the complex wave number plane. 
Appendix~\ref{appb} presents results obtained from a variational calculus.

%%%%%%%%%%%%%%%%%%%%%%%%%%%%%%%%%%%%%%%%%%%%%%%%%%%%%%%%%%%%%%%%%%%%
% \input{ii2703cj.tex}
%%%%%%%%%%%%%%%%%%%%%%%%%%%%%%%%%%%%%%%%%%%%%%%%%%%%%%%%%%%%%%%%%%%%

\section{SYSTEM}\label{ii}
We consider a horizontal layer of height $d$ of a binary
fluid mixture in the homogeneous gravitational field
${\bf g} = -g {\bf e}_z$ that is directed downwards.
A positive temperature difference $\Delta T$ is imposed
between the lower and upper confining boundaries, e.g.,
via highly conducting plates in experiments. 
The associated Rayleigh number is
\begin{equation}
Ra = \frac{\alpha g d^3}{\kappa \nu}\, \Delta T
\end{equation}
where $\kappa$ is the thermal diffusivity and $\nu$ the
kinematic viscosity.
The thermal expansion coefficient $\alpha$ and the solutal
expansion coefficient $\beta$ follow from a linear
isobaric equation of state for the total mass density
\begin{equation}
\rho = \rho_0\;\left[ 1-\alpha (T-T_0)-\beta (C-C_0)\right]
\end{equation}
for small deviations of the temperature $T$ from its 
mean $T_0$ and small deviations of the solute's mass 
concentration $C$ from its mean $C_0$.

An externally applied lateral pressure gradient drives 
a through-flow in $x$ direction. The resulting mean lateral
flow velocity $\overline{U}$ determines the
through-flow  Reynolds number
\begin{equation}
Re = \overline{U} \frac{d}{\nu}.
\end{equation}
We investigate here the parameter regime 
$ 0 \le Re \le 1$. With $d \approx 0.5~\mbox{cm}$, $\nu \approx
0.01~\mbox{cm}^2/s$ (H$_2$O) the maximal averaged through-flow velocity
is then $\overline{U} \approx 0.02~\mbox{cm/s}$, i.e., $1.2~\mbox{cm}$
per minute.

\subsection{Equations}\label{ii.a}
To describe this system we use the balance equations for 
mass, momentum, heat, and concentration in Oberbeck-Boussinesq 
approximation \cite{pl84,ch68}
\begin{mathletters}
\label{eqs}
\begin{eqnarray}
{\bf \mbox{\boldmath $\nabla$}\cdot u} &=& 0 \label{4a} \\
(\,\partial_t + {\bf u\cdot\mbox{\boldmath $\nabla$}}\,)\,{\bf u}& = &
\sigma\, {\bf \nabla}^2 \, {\bf u} \, - \mbox{\boldmath $\nabla$} p \nonumber \\
 & & \quad + \; \sigma \,(\,T - T_0\,+\,C - C_0) \,{\bf e_z}
\label{eqsa}  \\
(\,\partial_t + {\bf  u\cdot\mbox{\boldmath $\nabla$}}\,)\,T & = &
\, {\bf \nabla}^2\,T \label{eqsb}  \\
(\,\partial_t + {\bf  u\cdot\mbox{\boldmath $\nabla$}}\,)\,C & = &
L\,{\bf \nabla}^2 \left(\,C - \psi\, T \right) \quad. \label{eqsc}
\end{eqnarray}
\end{mathletters}
Here ${\bf u} = u\, {\bf e}_x + v\, {\bf e}_y + w\, {\bf e}_z $ is
the velocity field. 
We reduce lengths by $d$, times by $d^2 / \kappa$,
the effective pressure $p$ by $\kappa^2 / d^2$, 
temperatures by $\kappa \nu / (\alpha g d^3)$, and
the concentration field by $\kappa \nu / (\beta g d^3)$.
Then the two material parameters Prandtl number
\begin{equation}
\sigma = \frac{\nu}{\kappa}
\end{equation}
and Lewis number 
\begin{equation}
L = \frac{D}{\kappa}
\end{equation}
appear with $D$ being the concentration diffusion constant. 
Furthermore, there enters the separation ratio
\begin{equation}
\psi = - \frac{\beta}{\alpha}\, \frac{k_T}{T_0}
\end{equation}
that measures the strength of the linear Soret coupling 
between concentration and temperature field via the
thermodiffusivity $k_T$.

\subsection{Conductive state}\label{ii.b}
For small $Ra,Re$ a laterally homogeneous solution of (\ref{eqs})
is stable that describes a conductive state without
vertical convective flow. It is a combination of plane
horizontal Poiseuille flow
\begin{mathletters}
\label{ucond}
\begin{eqnarray}
{\bf u}_{cond} &=& U(z)\; {\bf e}_x = \sigma\, Re\, 
P(z)\, {\bf e}_x  \label{uconda} \\
P(z) &=& 6\, z\, ( 1 - z ) \label{ucondb}
\end{eqnarray}
\end{mathletters}
and a diffusive temperature field
\begin{equation}
T_{cond} = T_0 + Ra\, (\frac{1}{2}-z) \label{tcond} 
\label{t_cond}
\end{equation}
that enforces via the Soret effect a diffusive vertical 
concentration stratification
\begin{equation}
C_{cond} = C_0 - Ra\, \psi\, (\frac{1}{2}-z) \label{ccond} \quad.
\label{c_cond}
\end{equation}

%%%%%%%%%%%%%%%%%%%%%%%%%%%%%%%%%%%%%%%%%%%%%%%%%%%%%%%%%%%%%%%%%%%%
% \input{iii2703cj.tex}
%%%%%%%%%%%%%%%%%%%%%%%%%%%%%%%%%%%%%%%%%%%%%%%%%%%%%%%%%%%%%%%%%%%%

\section{CONVECTIVE PERTURBATIONS}\label{iii}
Here we briefly review the linearized equations for 
perturbations of the conductive state, their
boundary conditions, the eigenvalue problem for the
characteristic exponents, relevant symmetry properties,
and the behavior for small through-flow rates.
\subsection{Linearization around the conductive state}\label{iii.a}
\subsubsection{Equations}
The basis for our linear analysis of convective
perturbations of the conductive state described in
Sec.~\ref{ii.b} are the linearized field equations
\begin{mathletters}
\label{alleqs}
\begin{eqnarray}
\left( \partial_t - \sigma {\bf \nabla}^2 \right)
{\bf \nabla}^2  w \, + & & \nonumber \\
\left(\, U {\bf \nabla}^2 - \partial_z^2 U\, \right)\, \partial_x w &=& 
\sigma \left( \partial_x^2 + \partial_y^2 \right)
\left( \theta + c \right) \label{alleqsa} \\ 
\left( \partial_t - {\bf \nabla}^2 + U \partial_x
\right) \theta &=& Ra\, w \label{alleqsb} \\
\left( \partial_t - L {\bf \nabla}^2 + U \partial_x
\right) c &=& \psi \,Ra\, w - L \psi {\bf \nabla}^2\theta  
\label{alleqsc}
\end{eqnarray}
\end{mathletters}
for the deviations
\begin{equation}
\theta = T - T_{cond},\quad c = C - C_{cond}
\end{equation}
from the conductive state (\ref{t_cond},\ref{c_cond}). 
Here $w$ is the vertical 
velocity field that vanishes in the conductive state.
To derive (\ref{alleqsa}) we have applied twice the curl operator
to equation (\ref{eqsa}) using (\ref{4a}).
Note that the Poiseuille flow profile $U(z)$ of the
conductive state enters into Eqs. (\ref{alleqs}) making them 
non-autonomous.

\subsubsection{NSI boundary conditions}
We consider the horizontal boundaries to be perfectly
heat conducting and rigid with no slip and vanishing vertical
concentration transport. These, so called, NSI (no-slip,
impermeable) conditions impose 
\begin{equation}
\theta = w = \partial_z w = \partial_z \zeta =
 0 \quad \mbox{at}\quad z=0,1 \quad.
\label{randbed}
\end{equation}
Here we have introduced the combined field
\begin{equation}
\zeta = c - \psi \theta \quad.
\end{equation}
Since the concentration current at the no-slip boundaries is 
purely diffusive the condition $\partial_z \zeta = 0$ ensures
impermeability of the horizontal boundaries. Laterally we assume
the system to be unbounded.

\subsubsection{FS boundary conditions}
As an illustrative special case let us consider for the moment 
free slip (FS) horizontal boundaries with a shear-free plug 
flow profile, $U_{FS}(z) = \overline{U}$. In this idealized 
situation the effect of through-flow in Eqs. (\ref{alleqs}) can be
transformed away by a Galilei transformation to a system
that comoves with the vertically constant plug flow velocity
$\overline{U}=\sigma Re$. Thus, the stability properties of the
FS conductive state are not changed by the horizontal plug flow.
The stationary and oscillatory marginal stability curves 
of the mixture remain the same, only the characteristic
exponents acquire an additional imaginary part of size $i k_x 
\overline{U}$. This holds for FS horizontal boundaries
irrespective whether they are permeable or impermeable to the
concentration field $c$.

\subsection{Eigenvalue problem}\label{iii.b}
The general solution of the perturbation equations (\ref{alleqs}) 
can be written as a superposition of plane-wave perturbations with
lateral wavevector
\begin{equation}
{\bf k} = k_x {\bf e}_x + k_y {\bf e}_y \quad.
\end{equation}
The plane-wave solution ansatz for the fields
\begin{equation}
\Phi = ( w, \theta, \zeta )
\end{equation}
reads
\begin{equation}
\Phi({\bf r},t) = \widehat{\Phi}(z)\; e^{i ( k_x x + k_y y)}\;
e^{s\,t} \label{ansatz}
\end{equation}
with a complex characteristic exponent
\begin{equation}
s = \Re s + i\, \Im s = \gamma - i \omega 
\end{equation}
and complex $z$-dependent amplitude functions $\widehat{\Phi}
= (\widehat{w},\widehat{\theta},\widehat{\zeta})$. Inserting the
ansatz (\ref{ansatz}) into the field equations (\ref{alleqs}) 
yields the $3 \times 3$ linear eigenvalue problem
\begin{mathletters}
\label{linpr}
\begin{equation}
\left( {\cal L} + s\,{\cal M} \right)\; \widehat{\Phi}(z) = 0
\label{linpra}
\end{equation}
for the eigenvalues $s$ and eigenvectors $\widehat{\Phi}$ with
\begin{equation}
{\cal L} = {\cal L}^{(0)} + i\, \sigma\, k_x\, Re\, {\cal L}^{(1)}
\label{linprb}
\end{equation}

\begin{equation}
{\cal L}^{(0)} = \left(
\begin{array}{ccc}
-\sigma \left( \partial_z^2 - k^2 \right)^2 & 
\sigma (1+\psi) k^2 & \sigma k^2 \\
-Ra & k^2 - \partial_z^2 & 0 \\
0 & \psi \left( \partial_z^2 - k^2 \right) & L 
\left( k^2 - \partial_z^2 \right)
\end{array} \right).
\end{equation}
Into
\begin{equation}
{\cal L}^{(1)} = \left(
\begin{array}{ccc}
P \left( \partial_z^2 - k^2 \right) - \partial_z^2 P & 0 & \quad 0\\
0 & P & \quad 0 \\
0 & 0 & \quad P
\end{array} \right)
\end{equation}
enters the vertical profile $P(z)$ (\ref{ucondb}) of the Poiseuille
through-flow and its second derivative.
Finally

\begin{equation}
{\cal M} = \left( 
\begin{array}{ccc}
\partial_z^2 - k^2  & 0 & \quad 0 \\
0 & 1 & \quad 0 \\
0 & 0 & \quad 1
\end{array} \right). 
\end{equation}
\end{mathletters}
In the absence of through-flow ${\cal L}$ reduces to ${\cal L}^{(0)}$.

Due to the boundary conditions the eigenvalue spectrum is discrete.
We are interested in the three characteristic exponents $s_j\;(j=1,2,3)$
whose growth rates $\gamma_j$ are closest to zero and whose  
eigenfunctions $\widehat{\Phi}_j(z)$ have no nodes other than at
the horizontal boundaries $z=0,1$.

\subsection{Symmetries}\label{iii.c}
The solution of (\ref{linpr}), i.e., eigenvalue $s$ and eigenfunction
$\widehat{\Phi}$ depend on the material parameters $\sigma$ and $L$,
the control parameters $\psi,~Ra,~Re,$ and on the lateral wavevector
${\bf k}$. Note first of all that the dynamics of perturbations with
wavevectors perpendicular to the through-flow are not
changed by the latter, since for them the contribution from ${\cal L}^{(1)}$
vanishes when $k_x = 0$.

\subsubsection{Squire transformation}
Since ${\bf k}$ and $Re$ enter into (\ref{linpr}) only as $k^2$ and $k_x Re$
the dependence of the functions $f=s,~\widehat{\Phi}$ on ${\bf k}$ and $Re$ is
\begin{equation}
f = f\left( k^2,\; k_x Re \right).
\end{equation}
Using this behavior the Squire transformation \cite{hs33}
\begin{mathletters}
\label{squire}
\begin{eqnarray}
f \left( k_x^2 + k_y^2,\; k_x Re \right) &=& 
f \left( \widetilde{k}_x^2, \widetilde{k}_x \widetilde{Re} \right) \\
\widetilde{k}_x^2 = k_x^2 + k_y^2 &;& \widetilde{Re} = 
\frac{k_x}{\widetilde{k}_x}\,Re
\end{eqnarray}
\end{mathletters}
relates the functions $f$ for a wavevector with arbitrary
components $k_x,~k_y$ to the functions that have been determined
for wavevectors $\widetilde{{\bf k}} = \widetilde{k}_x {\bf e}_x$ in
through-flow direction and Reynolds numbers $\widetilde{Re}$.
Therefore, we shall consider in the remainder of this work 
only $k_y = 0$ perturbations with wavevectors ${\bf k}=k_x {\bf e}_x$
that are parallel or antiparallel to the through-flow.
For $Re=0$ the Squire relations (\ref{squire}) reflect the horizontal
rotational symmetry of the system in the absence of through-flow.

\subsubsection{Reverting the through-flow direction}
Upon reverting the flow direction, i.e., under the operation
$Re \rightarrow -Re$ the set $\{ s_j,\; \widehat{\Phi}_j \}$
of eigenvalues and eigenfunctions transforms into each other
since the balance equations (\ref{eqs}) are invariant under the parity
operation $(x,u) \rightarrow -(x,u)$ with $u$ being the velocity field
in $x$-direction. The transformation
behavior of $\{ s_j,\; \widehat{\Phi}_j \}$ follows explicitly from 
the fact that the linear 
operator ${\cal L}$ entering the eigenvalue equation (\ref{linpr})
transforms as
\begin{equation}
{\cal L}( -q_x ) = {\cal L}^* ( q_x ) \quad ;\quad q_x=k_x Re \label{traf1}
\end{equation}
under $Re \rightarrow -Re$ with the star denoting complex conjugation.
Here we do not display the other
arguments of ${\cal L}$ that remain unchanged. We use (\ref{traf1})
in the complex conjugate of Eq. (\ref{linpr})
\begin{equation}
\left[ {\cal L}(-q_x) + s^*(q_x)\,{\cal M} \right]
\; \widehat{\Phi}_j^* (q_x) = 0 \quad.
\end{equation}
Thus, if $s_j(q_x)$ with $\widehat{\Phi}_j(q_x)$ are
solutions of (\ref{linpr}) so are
\begin{equation}
\widetilde{s}_j(q_x) = s_j^*(-q_x) \quad \mbox{with}  \quad 
\widetilde{\widehat{\Phi}}_j(q_x) = \widehat{\Phi}_j^*(-q_x).
\label{traf2}
\end{equation}
Now, the nondegeneracy of the eigenvalue problem implies that
the two sets $\{ \widetilde{s}_j, \widetilde{\widehat{\Phi}}_j \}$ and
$\{s_j, \widehat{\Phi}_j \}$ are the same. We find that one
eigenvalue, say $j=3$, does not change --- $\widetilde{s}_3=s_3$
and $\widetilde{\widehat{\Phi}}_3 = \widehat{\Phi}_3$ --- so that
according (\ref{traf2})
\begin{mathletters}
\label{rel1}
\begin{eqnarray}
\gamma_3(-q_x) &=& \gamma_3(q_x) \\
-\omega_3(-q_x) &=& \omega_3(q_x) \\
\widehat{\Phi}_3^*(-q_x) &=& \widehat{\Phi}_3(q_x) \quad.
\end{eqnarray}
\end{mathletters}
The other two eigenvalues and eigenfunctions are crossrelated to
each other --- $\widetilde{s}_1 = s_2,\; \widetilde{\widehat{\Phi}}_1 =
\widehat{\Phi}_2$ and $\widetilde{s}_2 = s_1,\;
\widetilde{\widehat{\Phi}}_2 = \widehat{\Phi}_1$ --- so that according
to (\ref{traf2})
\begin{mathletters}
\label{rel2}
\begin{eqnarray}
\gamma_1(-q_x) &=& \gamma_2(q_x) \\
-\omega_1(-q_x) &=& \omega_2(q_x) \\
\widehat{\Phi}_1^*(-q_x) &=& \widehat{\Phi}_2(q_x) \quad.
\end{eqnarray}
\end{mathletters}
For $Re=0$ the relations (\ref{rel1}, \ref{rel2}) reflect 
the facts that in the
absence of through-flow ({\it i}) the solutions depend on $k_x^2$
only, ({\it ii}) one eigenvalue is real, say $\omega_3=0$, with 
real eigenfunction $\widehat{\Phi}_3=\widehat{\Phi}^*_3$ that
describes nonoscillatory dynamics, and ({\it iii}) the other two
are a complex conjugate pair --- $\gamma_1=\gamma_2,\; 
\omega_1=-\omega_2$ with $\widehat{\Phi}_1^*=\widehat{\Phi}_2$ ---
that describes the growth/decay of symmetry degenerate left and
right traveling waves. This symmetry is broken by the through-flow.
If for $Re=0$ all three eigenvalues and eigenfunctions are real
then they fulfill for finite $Re$ relations like (\ref{rel1}).

Since (\ref{rel1}, \ref{rel2}) relate the solutions 
for negative $k_x$ or $Re$ to those with 
positive $k_x,\; Re$ it suffices to investigate  
the three eigenvalues $s_j$ and eigenfunctions 
$\widehat{\Phi}_j$ for positive $k_x,\, Re$ only.

\subsection{Expansion for small Reynolds numbers}\label{iii.d}
It is instructive to see how the $Re=0$ solutions 
$\{ s_j, \widehat{\Phi}_j\}$
of the eigenvalue problem (\ref{linpr}) evolve upon switching on
the through-flow. The qualitative behavior can be 
studied {\it analytically} for small $Re$ via 
an expansion in the parameter
\begin{equation}
\eta = \sigma q_x = \sigma\, k_x\, Re
\end{equation}
that appears explicitly in the linear operator 
${\cal L}$ (\ref{linprb})
\begin{mathletters}
\label{entw}
\begin{equation}
{\cal L} = {\cal L}^{(0)} + i\, \eta\, {\cal L}^{(1)} \quad.
\end{equation}
So we expand
\begin{eqnarray}
s_j &=& s_j^{(0)} + \eta\, s_j^{(1)} + O(\eta^2) \\
\widehat{\Phi}_j &=& \widehat{\Phi}_j^{(0)} + 
\eta\, \widehat{\Phi}_j^{(1)}
+ O(\eta^2) \quad.
\end{eqnarray}
\end{mathletters}
Inserting (\ref{entw}) into (\ref{linpra}) yields in 
order $\eta$ the equation
\begin{equation}
\left( i\, {\cal L}^{(1)} + s_j^{(1)} 
{\cal M} \right) \widehat{\Phi}_j^{(0)}
= - \left( {\cal L}^{(0)} + s_j^{(0)} 
{\cal M} \right) \widehat{\Phi}_j^{(1)} 
\end{equation}
which is solvable under the condition
\begin{eqnarray}
\left< \widehat{\Phi}_j^{(0)\, \dagger} 
\left| \left( i\, {\cal L}^{(1)} + s_j^{(1)} {\cal M} \right) 
\widehat{\Phi}_j^{(0)} \right> \right. &=&  \nonumber \\ 
\int\limits^1_0 dz \left( \widehat{\Phi}_j^{(0)\, \dagger} \right)^* \; 
\left( i\, {\cal L}^{(1)} + s_j^{(1)} {\cal M} \right)
\widehat{\Phi}_j^{(0)} &=& 0 \label{fred} \quad.
\end{eqnarray}
Here $\widehat{\Phi}_j^{(0)\, \dagger}$ is the solution of the 
adjoint equation 
\begin{equation}
\left( {\cal L}^{(0)\,\dagger} + s_j^{(0)\,\dagger} {\cal M}^\dagger \right)
\widehat{\Phi}_j^{(0)\, \dagger} = 0 
\end{equation}
of the zeroth-order eigenvalue problem
\begin{equation}
\left( {\cal L}^{(0)} + s_j^{(0)} {\cal M} \right)
\widehat{\Phi}_j^{(0)} = 0 
\end{equation}
for the eigenvalue $s_j^{(0)}$. Note that  $s_j^{(0)\,\dagger}
= s_j^{(0)\,*}$ is the complex conjugate of the original eigenvalue
 $s_j^{(0)}$.
The Fredholm alternative (\ref{fred}) leads to the first-order
correction for $s_j$
\begin{equation}
s_j^{(1)} = -i\, p_j
\end{equation}
where
\begin{equation}
p_j =  \frac {\left< \;
\widehat{\Phi}_j^{(0)\,\dagger} \; \left| \;
{\cal L}^{(1)}\,
\widehat{\Phi}_j^{(0)} \right. \; \right>}
{ \left< \; \widehat{\Phi}_j^{(0)\,\dagger} \; \left| \;
{\cal M}\,
\widehat{\Phi}_j^{(0)} \right. \; \right>} 
\label{matrix}
\end{equation}
is determined by normalized "matrix elements" of through-flow
perturbation "operators" containing the profile $P(z)$ between
the zeroth-order eigenfunctions $\widehat{\Phi}_j^{(0)} = \left(
\widehat{w}_j^{(0)} , \widehat{\theta}_j^{(0)} , 
\widehat{\zeta }_j^{(0)} \right)$.
Thus, the small-$Re$ expansion yields the following results for
the eigenvalues $s_j=\gamma_j - i\, \omega_j$
\begin{mathletters}
\label{eig_entw}
\begin{eqnarray}
\gamma_j &=& \gamma_j^{(0)} + \eta\, \Im p_j + O(\eta^2) \\
\omega_j &=& \omega_j^{(0)} + \eta\, \Re p_j + O(\eta^2) \quad.
\end{eqnarray}
\end{mathletters}

For the subsequent discussion we use the fact that the "operators"
entering into the "matrix elements" of Eq. (\ref{matrix}) are real.

\subsubsection{Stationary perturbations}
Let us call a perturbation stationary, for shorthand, if the
characteristic exponent $s_j^{(0)}$ in the absence of through-flow
is real. Then the corresponding eigenfunction is real as well,
$\widehat{\Phi}_j^{(0)} 
= \widehat{\Phi}_j^{(0)\,*}$, which implies that also $p_j$ 
(\ref{matrix}) is real. Thus, with $\Im p_j = 0$, one obtains
from (\ref{eig_entw})
\begin{mathletters}
\label{eig_entw_st}
\begin{eqnarray}
\gamma_j &=& \gamma_j^{(0)} + O(Re^2) \\
\omega_j &=& \omega_j^{(0)} + \sigma\, k_x\, Re\, \Re p_j +
O(Re^2) \quad.
\end{eqnarray}
\end{mathletters}
The frequency grows linearly with $k_x Re$ while the growth 
rate is an even function of $k_x Re$ for these stationary
perturbations as shown schematically in Fig.~\ref{skizze1} 
for the eigenvalue labelled by $j=3$.

\subsubsection{Oscillatory perturbations}
We call the perturbations oscillatory that are described by the
two eigenvalues (say, $j=1,2$) which in the absence of through-flow
form a complex conjugate pair, $s_2^{(0)}=s_1^{(0)\,*}$.
The corresponding eigenfunctions are complex conjugates of
each other,  $\widehat{\Phi}_2^{(0)}=\widehat{\Phi}_1^{(0)\,*}$,
according to (\ref{rel2}) which implies $p_2 = p_1^*$.
With $\Re p_2 = \Re p_1$ and $\Im p_2 = -\Im p_1$ one obtains
from (\ref{eig_entw}) the following relations
\begin{mathletters}
\label{eig_entw_os}
\begin{eqnarray}
\gamma_{1 \atop 2} & = &
\quad \gamma_1^{(0)}  \pm  
\sigma k_x Re\, \Im p_1
 +  O(Re^2) \quad \label{eig_entw_os_a} \\
\omega_{1 \atop 2} & = &\, \pm \omega_1^{(0)}
 +  \sigma k_x Re\, \Re p_1  + 
O(Re^2) \quad. \label{eig_entw_os_b} 
\end{eqnarray}
\end{mathletters}
Since with the plane-wave perturbation ansatz (\ref{ansatz}) the TW
phase velocity $\omega/k_x$ should be increased for positive $k_x,~Re$
by the through-flow one can expect without having performed an explicit
calculation that $\Re p_1$ in (\ref{eig_entw_os_b}) should be
positive --- cf. Fig.~\ref{skizze1}. And consequently both
frequencies $\omega_{1,2}^{(0)}$ are shifted upwards by the same amount
$ \sigma\, k_x\, Re\, \Re p_1 $. On the other hand, to decide whether the
growth rate of the left or of the right traveling wave is 
increased or decreased requires an explicit calculation of $\Im p_1$.
In any case, however, Eq. (\ref{eig_entw_os_a}) predicts that for
small through-flow the symmetry degeneracy, $\gamma_1^{(0)} = \gamma_2^{(0)}$, 
of the growth rates is lifted by a symmetric splitting
that increases linearly with $Re$ as indicated schematically in Fig.~\ref{skizze1} 
for $\gamma_1$ and $\gamma_2$.

For idealized FS boundary conditions the frequencies
$\omega_j = \omega_j^{(0)} + \sigma\, k_x\, Re$ 
behave as in Fig.~\ref{skizze1}. However, the growth rates are
independent of $Re$ so that the degeneracy 
$\gamma_1^{(0)} = \gamma_2^{(0)}$ is not lifted by the FS plug flow.

%%%%%%%%%%%%%%%%%%%%%%%%%%%%%%%%%%%%%%%%%%%%%%%%%%%%%%%%%%%%%%%%%%%%
% \input{iv1104ml.tex}
%%%%%%%%%%%%%%%%%%%%%%%%%%%%%%%%%%%%%%%%%%%%%%%%%%%%%%%%%%%%%%%%%%%%

\section{STABILITY AND BIFURCATION PROPERTIES}\label{iv}
Here were present for negative as well as for positive
Soret coupling $\psi$ the $Re$-dependence of critical 
properties: stability thresholds, wave numbers, 
frequencies, and eigenfunctions. These results have
been obtained numerically by a variant \cite{cj93}
(cf. Appendix~\ref{appa.1} for a short description)
of a standard shooting method that has
previously been used \cite{wh88} to determine stability 
properties of binary mixtures in the absence of 
through-flow \cite{h92}. In order to check these
results --- in particular some of the unexpectedly
strong and peculiar changes with $Re$ --- by an
independent method we have performed a 
variational calculation described in Appendix~\ref{appb}.

\subsection{Notation}\label{iv.a}
We introduce Rayleigh numbers and wave numbers
\begin{mathletters}
\begin{eqnarray}
r &=& \frac{Ra}{Ra_c(Re=0,\psi=0)} \\  
\widehat{k} &=& \frac{k}{k_c(Re=0,\psi=0)}
\end{eqnarray}
\end{mathletters}
that are reduced by the critical ones of a pure fluid
($\psi=0$) in the absence of flow ($Re=0$)
\begin{mathletters}
\begin{eqnarray}
Ra_c(Re=0,\psi=0) &=& 1707.76 \\
k_c(Re=0,\psi=0) &=& 3.11632 \quad.
\end{eqnarray}
\end{mathletters}
As explained in Sec.~\ref{iii.c}, it suffices to consider
wave vectors ${\bf k} = k_x\, {\bf e}_x$ in through-flow
direction with positive component $k_x = | {\bf k} | = k$.
The spatiotemporal behavior of perturbations with other vectors
follows with the symmetries of Sec.~\ref{iii.c}. 
We determine the evolution of the three relevant eigenvalues
$ s_j = \gamma_j - i\, \omega_j$ (cf. Sec.~\ref{iii.b}) upon 
increasing $Re$ from $Re=0$ up to about $1$. In particular
we evaluate the critical parameter combinations for which
each of the three growth rates $\gamma_j$ first passes through zero
when increasing $r$.

The so obtained critical quantities are marked by
a subscript $c$ and in addition by a superscript $S,~U,~D$
that replaces the running index $j$ of the three eigenvalues
used in Sec.~\ref{iii}. The superscripts $S$, $U$, and $D$ identify the 
critical perturbation behavior, $e^{i\,(k_c\,x - \omega_c\,t)}$,
in the limit $Re \rightarrow 0$. Eigenvalues for which
$\omega_c(Re \rightarrow 0) = 0$ are marked by $S$ since these
perturbations are stationary for $Re=0$. Eigenvalues for which
$\omega_c(Re \rightarrow 0)$ is positive (negative) carry the
superscript $D$ ($U$) since they characterize for  $Re \rightarrow 0$
perturbations which propagate in downstream (upstream) direction.
We should like to stress again that the cases $S$ ("stationary"),
$D$ ("downstream"), and $U$ ("upstream") characterize the 
perturbations in the limit $Re \rightarrow 0$ 
--- see also Fig.~\ref{skizze1}. In general all critical
frequencies are finite in the presence of through-flow.
However, for a special value of $Re$ one has $\omega_c^U = 0$
while  $\omega_c^S$ and  $\omega_c^D$ are positive 
(cf. Sec.~\ref{iv.b} and Sec.~\ref{iv.c}).

\subsection{Effect of through-flow on the oscillatory instability}\label{iv.b}
We present in this subsection critical properties
of a binary mixture like water-ethanol with $L=0.01,\;\sigma=10$,
and separation ratio $\psi=-0.1$ as a representative case for
a moderately negative Soret coupling. For these parameters the 
nonlinear solution of stationary convection in the absence
of through-flow, $Re=0$, is already disconnected from the 
conductive state since the stationary bifurcation threshold
$r_c^S$ has moved already at $\psi_{\infty}^0=-L/(1+L)$ to infinity
\cite{LLT83}. However, at $r_c^D = r_c^U = 1.1200$
there is for $Re=0$ an oscillatory threshold into symmetry
degenerated left and right --- or in our language upstream and
downstream --- propagating traveling waves with critical
wave numbers $\widehat{k}_c^D = \widehat{k}_c^U = 1.0022$
and critical Hopf frequency $\omega_c^D = -\omega_c^U = 6.4659$.

In Fig.~\ref{krw_01} we show the variation of these two critical
thresholds, wave numbers, and frequencies with increasing through-flow
Reynolds numbers. In each case the full (dashed) line represents the
upstream (downstream) critical quantity. Initially, for small $Re$,
the bifurcation threshold $r_c^U (r_c^D)$ is depressed (enhanced) by 
the through-flow so that the conductive state is destabilized 
(stabilized) against convective perturbations traveling upstream 
(downstream) in comparison to the symmetry degenerate Hopf bifurcation
threshold without through-flow --- see also the schematic variation of
the growth rates in Fig.~\ref{skizze1}. So at very small $Re$ it is 
upstream traveling wave convection that grows first when increasing the 
Rayleigh number quasistatically.
 
However, while $r_c^D$ increases monotonically with $Re$ --- first
linearly and then quadratically --- the initial linear downwards shift
of $r_c^U$ changes at larger $Re$ to a precipitous increase and a 
subsequent flattening. Thus, in the shown $Re$-range the two critical
curves $r_c^U$ and $r_c^D$ have two intersections giving rise to 
bistable behavior of perturbations there. Note, however, that the 
bistable upstream and downstream TW perturbations can not be superimposed
linearly to a standing wave since their wave numbers $k_c^U \neq k_c^D$
differ and furthermore $\omega_c^U \neq -\omega_c^D$. So in the 
$Re$-interval between the bistable intersections  of $r_c^U$ and $r_c^D$
downstream propagating convection waves grow first while outside this
interval at small $Re$ and large $Re$ upstream TW convection
bifurcates first out of the conductive state.

Considering the critical wave numbers $k_c^U$ and $k_c^D$ of Fig.~\ref{krw_01}b 
it should be noted that their variation is very small
--- less than $2\%$ --- and that $k_c^U < k_c^D$ in the $Re$-range of 
Fig.~\ref{krw_01}. This behavior holds also for the other Soret coupling 
strengths $\psi = -0.25, -0.01, \mbox{and}\, -0.001$ that we have
investigated \cite{cj93}.

Somewhat unexpected to us is the non monotonical variation of $k_c^U$
and also of $\partial \omega_c^U / \partial Re$ (Fig.~\ref{krw_01}c) in the
interval below $Re=0.5$ where $\omega_c^U$ changes sign and where
$r_c^U$ shows its strong increase. To check that this variation
is not a numerical artifact of our shooting algorithm
we have performed a stability analysis with a variational approximation
being a fundamentally different method. The variational results
presented in Appendix~\ref{appb} also show the peculiar variation of $k_c^U$
with $Re$ obtained from the shooting method thus supporting the latter
behavior.

The critical frequencies $\omega_c^U$ and $\omega_c^D$ shown in
Fig.~\ref{krw_01}c are practically linear functions of $Re$ and in this 
respect similar to the frequencies of the idealized FSP system.
They start at zero through-flow with the Hopf values $|\omega_c^{(0)}|$
and $-|\omega_c^{(0)}|$, respectively, and they
can be very well approximated by the first-order result (\ref{eig_entw_os_b})
of the low-$Re$ expansion
\begin{equation}
\omega_c = \omega_c^{(0)} + \sigma\,k_c^{(0)}\, \Re p_1\, Re
\label{om_appr}
\end{equation}
with $\sigma\,k_c^{(0)}\, \Re p_1 \approx 41.9$ for $\sigma=10$. 
Comparing this rate
of change $\partial \omega_c / \partial Re \approx 41.9$ with results 
for other separation ratios including the pure fluid case \cite{wm92} 
one finds only very small deviations \cite{cj93}. Obviously $\Re\, p_1$
(\ref{matrix}) depends only weakly on the Soret coupling $\psi$.

Note that for $Re \ge |\omega_c^{(0)}|/41.9$ 
both critical frequencies, $\omega_c^D$ and $\omega_c^U$, are 
positive. Then the phase velocities, $v_c=\omega_c / k_c$, of 
the two different critical TW`s are in 
positive $x$-direction in the laboratory system, i.e., in 
through-flow direction. However, 
$v_c^U$ is always smaller --- by about 
$2|v_c^{(0)}|$ --- than $v_c^D$.
Only for $Re \le |\omega_c^{(0)}|/41.9$ is the phase 
velocity $v_c^U$ negative, i.e., opposite to the 
through-flow. So the wording 
"upstream propagating perturbations" that we are 
using in this work does not 
necessarily imply that the phase velocity 
of such a TW is negative in the 
laboratory frame.
It would be negative in a frame 
that is moving in through-flow direction with 
a conveniently defined mean lateral velocity like, e.g.,
$\overline{v} = \frac{1}{2} ( v_c^D + v_c^U )$.

\subsection{Bifurcation thresholds at negative $\psi$}\label{iv.c}
In Fig.~\ref{r_psi} we show the bifurcation thresholds 
$r_c^U$ (full lines), $r_c^D$ (dashed lines), and 
$r_c^S$ (dotted lines) as functions of $Re$ for a few 
characteristic negative Soret couplings $\psi$.

\subsubsection{$r_c^U(Re,\psi)$}
The typical shape of the stability curve $r_c^U$ that is 
displayed in Fig.~\ref{krw_01}a for $\psi=-0.1$ does not change much 
for other negative separation ratios: 
As a function of $Re$ $r_c^U$ (full lines in Fig.~\ref{r_psi}) 
decreases for small $Re$, develops a minimum where $\omega_c^U$ goes 
through zero, steeply increases thereafter, and 
finally flattens asymptotically towards $r_c^S(Re,\psi=0)$ at large $Re$.
Thus a sufficiently large through-flow eliminates the Soret induced coupling
effects between concentration field on one side and temperature and
velocity field on the other side: The bifurcation threshold $r_c^U(Re,\psi)$
approaches for any Soret coupling $\psi$ the pure fluid stability
boundary $r_c^S(Re,\psi=0)$ at large $Re$.
For small $\psi$, e.g., at $\psi=-0.001$, the stability boundary 
$r_c^U$ lies always below $r_c^D$ while for larger 
$|\psi|$  (see, e.g., $\psi=-0.01$) there are two intersections 
of the curves $r_c^D$ and $r_c^U$ with  $r_c^D \le r_c^U$ in between 
--- cf. the related discussion in Sec.~\ref{iv.b}.

\subsubsection{$r_c^D(Re,\psi)$}
The bifurcation threshold $r_c^D$ (dashed lines in 
Fig.~\ref{r_psi}) always 
increases monotonically with the through-flow strength.
The initial slope $\partial r_c^D / \partial Re $ increases 
somewhat with decreasing $|\psi|$. For $\psi=-0.001$ and 
$\psi=-0.01$ the stability curves $r_c^D$ and
$r_c^S$ collide in the $Re$-range displayed in Fig.~\ref{r_psi}. This 
property is elucidated in paragraph 4 further below.

\subsubsection{$r_c^S(Re,\psi)$}
In pure fluids, $\psi=0$, the bifurcation 
threshold $r_c^S(Re,\psi=0)$ (lowest dotted curve in 
Fig.~\ref{r_psi}) slightly 
increases with growing $Re$ \cite{wm92}. 
In binary mixtures with negative Soret coupling, 
on the other hand, the stationary 
threshold $r_c^S$ gets very strongly depressed by a small through-flow  
--- see the dotted curve for $\psi=-0.001$ that 
starts at $r_c^S(Re=0,\psi=-0.001)=1.1816$.

In the absence of through-flow, $Re=0$, the threshold $r_c^S$ rapidly 
increases with $|\psi|$ and diverges at 
$\psi_{\infty}^0=-L/(1+L)=0.0099$ for $L=0.01$. 
Beyond this Soret coupling the solution branch of 
stationary nonlinear convection 
is disconnected from the ground state solution 
as $r_c^S(Re=0,\psi \le \psi_{\infty}^0)=\infty$.
A small but finite through-flow, however, 
moves the threshold $r_c^S$ down to finite values: 
The dotted curve for $r_c^S$ in the inset of Fig.~\ref{r_psi} for 
$\psi=-0.01 < \psi_{\infty}^0$ shows {\it(i)} that 
$r_c^S=\infty$ below a finite $Re_{\infty} \approx 0.019$, 
{\it (ii)} that $r_c^S$ is finite for $Re > Re_{\infty}$, and {\it
(iii)} that $r_c^S$ 
steeply drops down for $Re > Re_{\infty}$.
The Reynolds number $Re_{\infty}$ where $r_c^S$ diverges 
grows with increasing $|\psi|$ --- a stronger Soret coupling
requires a larger through-flow to move the bifurcation threshold 
$r_c^S$ from infinity to a finite value.

\subsubsection{Collision of the $r_c^D$ and $r_c^S$
 stability boundaries}\label{iv.c.4}
With increasing $Re$ the bifurcation threshold $r_c^S$ and 
$r_c^D$ approach each other. The former decreases 
rapidly and the latter increases with $Re$ and they almost coalesce 
in the $Re-r$ plane of Fig.~\ref{r_psi}. This behavior is most easily understood 
by investigating how the relevant eigenvalues $s=\gamma - i\, \omega$ vary 
with $r$ and $Re$.
To that end we show in Fig.~\ref{eigw} $\gamma(r)$ 
and $\omega(r)$ of the two relevant eigenvalues for a representative Soret 
coupling  $\psi=-0.01$ at $Re=0$ (thick curves) 
and at $Re=0.4$ (thin curves) for a fixed wave number 
$\widehat{k}=1$. The real part of the third eigenvalue is always negative 
in the parameter range of Fig.~\ref{eigw} and thus 
irrelevant for the following. 
Fig.~\ref{eigw2} shows in a {\it schematic} way the motion of these 
two eigenvalues in the complex $s$-plane with increasing $r$ 
for $Re=0$ (thick curves) and for a small $Re\neq0$ (thin curves).

Let us consider first $Re=0$.
Then there is for small $r$ a complex conjugate pair 
of eigenvalues ($\gamma^U=\gamma^D, \omega^U=-\omega^D$) that 
produce the symmetry degenerate Hopf bifurcation 
(at $r^D=r^U \approx 1.021$ in Fig.~\ref{eigw}b) when $\gamma^U$ and 
$\gamma^D$ pass simultaneously through zero. This situation is 
marked in Fig.~\ref{eigw} and Fig.~\ref{eigw2} by thick upwards and 
downwards pointing triangles. 
With increasing $r$ the real parts $\gamma$ grow and the frequencies 
$\omega$ approach zero and the pair of eigenvalues 
meets in the complex $s$-plane of Fig.~\ref{eigw2} on the 
real $\gamma$-axis (i.e. in Fig.~\ref{eigw} at $r \approx 1.231$).
Then, at larger $r$, this pair of real eigenvalues splits and 
moves apart along the real $\gamma$-axis (cf. thickly dotted $S1$ and $S2$
curves in Fig.~\ref{eigw} and \ref{eigw2}).
So we have a transformation of two oscillatory eigenvalues 
($U$ and $D$) into two stationary ones ($S1$ and $S2$).
For $Re=0$ the two thickly dotted branches $\gamma^{S1}$ and $\gamma^{S2}$
in Fig.~\ref{eigw}b remain above zero, i.e., $S1$ does not reach 
the imaginary axis in Fig.~\ref{eigw2}.

Now for finite $Re$ the symmetry degeneracy of the 
Hopf eigenvalue pair is lifted. 
The $r$-value where $\gamma^U$ goes through zero 
($\bigtriangleup$ at $r^U \approx 1.05$ in Fig.~\ref{eigw}b) 
differs from the one where $\gamma^D=0$ 
($\bigtriangledown$ at $r^D \approx 1.112$ in Fig.~\ref{eigw}b)
and  the frequencies $\omega^U$ and $\omega^D$ (thin lines in 
Fig.~\ref{eigw}a) are shifted 
upwards. Thus the $Re=0$ pitchfork topology of the eigenvalue paths 
in Fig.~\ref{eigw2} 
is perturbed. Moreover, the eigenvalue branches become 
disconnected (thin lines in Fig.~\ref{eigw2}) as the thick eigenvalue 
branches of Fig.~\ref{eigw} are deformed 
by the trough-flow into the thin ones --- the arrows in 
Fig.~\ref{eigw}b indicate 
the deformation directions.
In particular the lower $\gamma^{S1}$ branch of 
Fig.~\ref{eigw}b and similarly the left moving 
$S1$ branch in Fig.~\ref{eigw2} 
goes at a sufficiently large $Re$
through zero at the open circle (${\large \circ}$) in 
Fig.~\ref{eigw} and \ref{eigw2} thereby 
producing an $S$-instability for $\widehat{k}=1$ at ${\large \circ}$, i.e., at a  
finite value of $r^S$ ($\approx 1.296$ in Fig.~\ref{eigw}). 
Coming from $\infty$ the intersection 
${\large \circ}$ in Fig.~\ref{eigw}b has moved
with increasing $Re$ to the left to the finite value $r^S$.
By increasing $Re$ further the thin $D/S1$-curve in Fig.~\ref{eigw}b is pushed 
downwards towards smaller $\gamma$, i.e., to the
left in  Fig.~\ref{eigw2}. Thereby the zero crossings at 
$r^D(\bigtriangledown)$ and $r^S({\large \circ})$ move together and vanish 
simultaneously at a particular $Re$-value. 
Thereafter there is only the zero crossing of $\gamma^U$ at 
$r^U(\bigtriangleup)$.
This is in principle what happens when the $r_c^D$ and  $r_c^S$ curves in 
Fig.~\ref{r_psi} at $\psi=-0.001$ and $\psi=-0.01$ approach each other
with increasing $Re$.

\subsubsection{Opening of a wave number gap in the $D-S$ marginal stability curves}\label{iv.c.5}
The above described merging of the zeros of the  $\gamma^{D/S1}$ 
curve that occurs in Fig.~\ref{eigw}b for $\widehat{k}=1$ slightly
above $Re=0.4$ corresponds to the opening up of a gap in the marginal
stability curves against $D$ and $S$ perturbations. This scenario is documented
in the $\widehat{k}-r$ plane of Fig.~\ref{OM}. There we show with gray-scales the height
distribution of $\gamma^{D/S1}$ in the $\widehat{k}-r$ range where $\gamma>0$.
Thick lines labelled by $\gamma=0$ are marginal stability curves. In the white
parts of Fig.~\ref{OM} $\gamma^{D/S1}$ is negative and the dashed
lines indicate $\gamma$-isolines for $\gamma<0$. The third eigenvalue 
$\gamma^{U/S2}$ being positive 
--- cf. Figs.~\ref{eigw} and ~\ref{eigw2} --- is not shown.
The open triangle and the circle in Fig.~\ref{OM}a mark the zeros
of $\gamma^{D/S1}$ at $Re=0.4$ that are shown in Fig.~\ref{eigw}b by the
same symbols.

Upon increasing $Re$ the eigenvalue $\gamma^{D/S1}$ decreases.
Thus, the "mountain landscape" of $\gamma^{D/S1}$ somewhat globally
"sinks" down. Thereby the $\gamma=0$ isolines come together in the
$\widehat{k}-r$ plane of Fig.~\ref{OM} and since the "mountain ridge" of 
$\gamma^{D/S1}$ does not have constant height (cf. gray-scales
in  Fig.~\ref{OM}a) the $\gamma=0$ curves are connected into two tongues (Fig.~\ref{OM}b)
that are separated by a wave number gap in which $\gamma^{D/S1}$
is negative. Increasing $Re$ further the gap widens and the $\widehat{k}-r$ 
regions with $\gamma>0$ between the $\gamma=0$ isolines narrow down as the latter
move away from $\widehat{k}=1$ towards larger $r$.

We have difficulties to resolve this behavior of the $\gamma=0$ isolines
with our shooting method. We therefore show in Fig.~\ref{r_psi} and later
in Figs.~\ref{3d} and \ref{konab} the minima $r_c^D$ and $r_c^S$ of
the marginal stability curves only {\it before} the wave number gap opens.
The ending of $r_c^D$ and $r_c^S$ in Figs.~\ref{r_psi}, ~\ref{3d}, and \ref{konab}
should therefore not be interpreted as a termination of bifurcation
branches: after the opening up of the gap the tongue shaped stability
curves move towards larger $r$ and with them their minima. 

\subsection{Bifurcation properties at positive $\psi$}\label{iv.d}
In the absence of through-flow there is only a stationary 
bifurcation threshold $r_c^S(Re=0,\psi)$ at $\psi\ge0$ that strongly 
drops from $r_c^S(Re=0,\psi=0)=1$ towards 
zero when increasing $\psi$. Switching on the through-flow has 
the overall effect of increasing $r_c^S$
towards $r_c^S(Re,\psi=0)$
as can be read off from Fig.~\ref{rwkpsi0}.
Thus the lateral flow stabilizes the basic state by 
eliminating in the so called Soret regime \cite{ms91}  the 
convectively induced concentration homogenization.
Note that already a very 
small through-flow has a dramatic stabilization effect: 
$r_c^S$ increases very strongly for small $Re$.
Similarly the critical wave number 
$\widehat{k}_c^S$ (Fig.~\ref{rwkpsi0}b)
approaches with increasing $Re$ the 
pure-fluid value $\widehat{k}_c^S(Re,\psi=0)$.

\subsection{Bifurcation surfaces in $r-Re-\psi$ space}\label{iv.e}
To give an impression of the form of the three critical surfaces
in the  $r-Re-\psi$ space where $U,~D,$ and $S$ convection patterns
bifurcate out of the conductive state we combine
in Fig.~\ref{3d} in a 3-D plot the $Re$-dependence of 
the bifurcation thresholds $r_c^U$ (thin full lines), $r_c^D$ 
(thin dashed lines), and 
$r_c^S$ (thin dotted lines) presented so far together with 
their $\psi$-dependence for $Re=0$ (thick lines).
The bifurcation surfaces $r_c^D(Re,\psi)$
and $r_c^U(Re,\psi)$ emanate for negative
Soret coupling $\psi<0$ out of the degenerate Hopf threshold
line $r_c^D(Re=0,\psi)=r_c^U(Re=0,\psi)$ (thick dashed and full 
line) and split apart when the through-flow is switched on.
Upon increasing $Re$ further $r_c^U$ gets indented slightly.
On the other hand,
$r_c^D$ curls up and comes very close to the surface $r_c^S(Re,\psi)$ that is
strongly bent down by the through-flow for $\psi<0$. See Sec.~\ref{iv.c.5}
for a discussion of the further fate of these bifurcation surfaces.

The physically relevant, i.e. lowest lying, 
surface at larger $Re$ is $r_c^U(Re,\psi)$ 
(thin full lines). For any $\psi<0$ it asymptotically approaches
with increasing through-flow  the $\psi=0$ stability threshold, i.e.,
$r_c^U(Re,\psi) \rightarrow r_c^S(Re,\psi=0)$. Thus a sufficiently
large through-flow effectively eliminates the influence of any
Soret coupling between {\it convective} concentration field and
temperature and velocity fields (the diffusively induced concentration
stratification in the conductive state (\ref{c_cond}), 
on the other hand, is not altered by the lateral shear flow ).
The bifurcation thresholds of the mixture approach that one,
$r_c^S(Re,\psi=0)$, of the pure fluid. This also holds for mixtures
with positive Soret coupling --- cf. the thin dotted line for
$r_c^S(Re,\psi)$ at $\psi>0$.     
 
\subsection{Structure of critical convective patterns}\label{iv.f}
Here we show how the spatial structure of the critical field
deviations from the conductive state changes with increasing $Re$.
To that end we have evaluated the complex eigenfunctions
$\Phi(x,z,t) = \widehat{\Phi}(z)\, e^{i\,(k_c x - \omega_c t)}$
at the critical thresholds $r_c^D$, $r_c^U$, and $r_c^S$.
Thus the critical convective fields $w, \theta, c$ have the form
\begin{equation}
w(x,z,t)=|\widehat{w}(z)|\;\cos \left[\, k_c x - \omega_c t - \varphi_w(z)\, \right]
\end{equation}
and similarly for $\theta$ and $c$. Being solutions of complex 
linear equations we choose the abitrary complex scaling constant
by fixing the modulus of the convective temperature field in the
middle of the layer,
\begin{equation}
|\widehat{\theta}(z=\frac{1}{2})| = 1 \;,
\end{equation}
and by fixing the vertical mean of the phase $\varphi_w(z)$ of the
vertical velocity field to zero
\begin{equation}
\int_0^1 dz\; \varphi_w(z) = 0.
\end{equation}

We present gray-scale contour plots of the fields  $w, \theta$, and $c$
in the vertical $x-z$ cross section of the fluid layer with white
(black) denoting large (small) values. Each of these figures~\ref{iso1},
\ref{iso2}, \ref{iso3} has nine contour lines denoting the fractions
$\pm n/5$ of the maximal field values with $n=0,1,2,3,4$.
In addition we present in figures~\ref{phase1},~\ref{phase2},~\ref{phase3}
vertical profiles of the moduli and phases of the complex field 
amplitudes $\widehat{w}, \widehat{\theta}$, and $\widehat{c}$.

\subsubsection{Propagating patterns for $Re=0$}
First we briefly recall the critical TW field structure
(top part of Fig.~\ref{iso1}) in the absence of through-flow
\cite{h92,wh88,km88,ck88-1} for a relatively large Soret coupling
$\psi=-0.25$. The TWs of velocity, temperature, and
concentration are vertically not plane but their phases show a
vertical variation that is largest for the concentration wave
(Figs.~\ref{iso1} and \ref{phase1}). The lateral location of a
concentration surplus (white ellipse in the $c$-field of Fig.~\ref{iso1})
phase-lags by about a quarter wavelength behind the lateral
position of vertical downflow (black ellipse in the $w$-field).
Thus, since the $w$-field advectively transports concentration
surplus (deficiency) from the Soret induced alcohol rich top
(poor bottom) boundary layer into the bulk fluid this feeding
mechanism between boundary layer and bulk is laterally phase shifted
by roughly $\lambda/4$ in the propagating wave. Also the crest (valley)
position of the temperature wave phase-lags --- albeit by a
smaller amount --- behind the lateral location of maximal vertical
upwards (downwards) flow which advectively feeds the bulk with
warm (cold) fluid from the warm bottom (cold top) region. This 
phase-lag agrees quite well with the result
\begin{equation}
\varphi_w - \varphi_{\theta} \approx \mbox{arctan}\, \left[
\frac{\omega_c^{(0)}}{\pi^2 + \left( k_c^{(0)} \right)^2} \right]
\end{equation}
obtained \cite{linz88} from a Galerkin model.

As a result of the
smallness of the Lewis number $L$, i.e., of the diffusive concentration
transport the $c$-field shows characteristic boundary layer behavior
near the plates where advection decreases to zero --- see the
variation of the modulus $|\widehat{c}(z)|$ and of the phase $\varphi_c(z)$
in Fig.~\ref{phase1}. Whithin the boundary layers $|\widehat{c}(z)|$
is suppressed and the phase-lag of the $c$-wave behind the $w$-wave
is significantly enlarged.

\subsubsection{"Downstream" patterns}
Upon turning on the through-flow the phaselines of the
downstream propagating patterns get bent further (cf. Fig.~\ref{iso1} and
in particular the right column of Fig.~\ref{phase1}) and the phase differences
between the different waves increase near the plates. The vertical
moduli profiles of $w$ and $\theta$ do not change. On the other hand,
$|\widehat{c}(z)|$ decreases (increases) near the plates (in the center
of the fluid layer) so that the vertical profile of $|\widehat{c}|$
flattens near the plates and becomes more peaked in the bulk near $z=1/2$.
The phase-lag of $\varphi_c(z)$ increases in the boundary layers with the
through-flow.

The above described flow-induced structural changes of $\widehat{c}(z)$
can easily be understood within the
$L=0$ approximation \cite{ck88-1,wh88} to the concentration field balance
(\ref{alleqsc})
\begin{equation}
\widehat{c}(z) \approx i\, \psi\, Ra_c^D\, 
\frac{\widehat{w}(z)}{\omega_c^D - \sigma\, k_c^D\, P(z)\, Re}
\label{l=0:1}
\end{equation}
in a critical TW that is propagating downstream with the critical
frequency $\omega_c^D$. Thus for $L=0$, i.e. in the absence of diffusive
concentration currents --- which, by the way, is quite a good approximation 
to the real situation of ethanol-water mixtures with $L=0.01$ --- the
$\widehat{c}(z)$-profile is advectively slaved to the vertical velocity
$\widehat{w}(z)$ and the through-flow velocity $U(z)=\sigma\,P(z)\,Re$:
The prefactor $i$ in (\ref{l=0:1}) reflects the overall phase shift of $\pi/2$
between $\widehat{w}$ and $\widehat{c}$. The {\it effective $z$-dependent}
lateral velocity 
\begin{equation}
u_{eff}^D(z) = v_c^D - \sigma\, P(z)\, Re
\end{equation}
that enters in the denominator of (\ref{l=0:1}) renormalizes the profile
$\widehat{w}$ in the numerator of (\ref{l=0:1}) so that the $L=0$
concentration profile
\begin{equation}
\widehat{c}(z) \approx i\, \psi\, \frac{Ra_c^D}{k_c^D}\, 
\frac{\widehat{w}(z)}{u_{eff}^D(z)}
\label{c_appr}
\end{equation}
is determined by the quotient of these two velocity profiles. Since
the phase velocity of the $D$-wave, 
$v_c^D = \omega_c^D / k_c^D$, can
well be approximated by
\begin{equation}
v_c^D \approx v_c^{(0)} + \sigma\, \Re p_1\, Re
\end{equation}
according to (\ref{om_appr}) with $\Re p_1 \approx 1.34$
we can rewrite the lateral velocity that is effective for the
concentration distribution profile (\ref{c_appr}) of the $D$-wave as
\begin{equation}
u_{eff}^D(z) \approx v_c^{(0)} + \sigma\, Re\, 
\left[ 1.34 - 6\,z\,(1-z)\, \right].
\end{equation}
Thus, in the bulk 
$u_{eff}^D(z=1/2)\approx v_c^{(0)}-0.16 \sigma Re$
decreases with increasing through-flow while $u_{eff}^D$ 
increases near the plates with $Re$, e.g., 
$u_{eff}^D(z=1/4) \approx v_c^{(0)} + 0.21 \sigma Re$. 
This explains the flow-induced changes in
the profile $\widehat{c}$ (\ref{c_appr}) to be seen in Fig.~\ref{phase1}
for $\psi=-0.25$ with $v_c^{(0)}=3.6$.

\subsubsection{"Upstream" patterns}
In the absence of through-flow the field structure of TWs propagating
to the left, i.e., in upstream direction is the mirror image of the
$Re=0$ downstream patterns shown in the top part of Fig.~\ref{iso1}.
Therefore we have not included the $Re=0$ reference upstream pattern in
Fig.~\ref{iso2} where we show upstream patterns for $Re=0.3, 0.45,$ and
$0.6$.

Switching on the through-flow decreases and eventually reverts the
original $Re=0$ phase bending of the $w$ and $\theta$ waves (cf. right
column in Fig.~\ref{phase2}) while the moduli $|\widehat{w}|$ and
$|\widehat{\theta}|$ remain practically unaffected by the through-flow of
Figs.~\ref{iso2} and \ref{phase2}. On the other hand, the upstream
concentration wave is significantly changed with increasing $Re$: 
$|\widehat{c}(z)|$ develops two side maxima in the upper and lower
half of the fluid layer while flattening in the center part
(cf. left column of Fig.~\ref{phase2} or also Fig.~\ref{iso2}).
These structural changes of the $c$-wave always occur in the $Re$-range
beyond the zero crossing of $\omega_c^U$ (Fig.~\ref{krw_01}) where
also $r_c^U$ and $k_c^U$ show a significant variation. Thus, we infer
that these phenomena are related to each other.

The flow-induced structural change of $\widehat{c}(z)$ can be
understood within the $L=0$ approximation
\begin{equation}
\widehat{c}(z) \approx i\, \psi\, \frac{Ra_c^U}{k_c^U}\, 
\frac{\widehat{w}(z)}{u_{eff}^U(z)}
\label{l=0:2}
\end{equation}
to the concentration balance (\ref{alleqsc}) in an upstream TW with effective
lateral velocity
\begin{equation}
u_{eff}^U(z) = v_c^U - \sigma\, P(z)\, Re \quad.
\end{equation}
Again $v_c^D = \omega_c^D / k_c^D$ can well be approximated by
\begin{equation}
v_c^U \approx -|v_c^{(0)}| + \sigma\, \Re p_1\, Re
\end{equation}
so that
\begin{equation}
u_{eff}^U(z) \approx -|v_c^{(0)}| + \sigma\, Re\, 
\left[ 1.34 - 6\,z\,(1-z)\, \right].
\label{l=0:3}
\end{equation}
Note, however, that ignoring the dissipative contribution $-L \nabla^2 c$
on the {\it lhs} of the concentration balance (\ref{alleqsc}) causes in the
$L=0$ concentration profile (\ref{l=0:2} - \ref{l=0:3}) a divergence at
\begin{mathletters}
\begin{equation}
z_{\pm} = \frac{1}{2} \pm \frac{1}{2} \sqrt{1 - \frac{2}{3}\, 
\left( 1.34 - \frac{|v_c^{(0)}|}{\sigma Re} \right)}
\end{equation}
via the zeros of $u_{eff}^U(z)$ whenever
\begin{equation}
Re >  \frac{|v_c^{(0)}|}{1.34 \sigma} \quad.
\end{equation}
\end{mathletters}
The diffusive contribution in (\ref{alleqsc}) prevents the divergence at
$z_{\pm}$ leading to side maxima in $|\widehat{c}(z)|$ instead (cf.
Fig.~\ref{phase2} for $Re=0.45$ and $0.6$). Their location is shifted
slightly towards the bulk in comparison with, e.g.,
$z_-(Re=0.45)=0.1$ and $z_-(Re=0.6)=0.144$.

It should be noted that for the large Soret coupling $\psi=-0.25$
the upstream fields diplayed in Fig.~\ref{iso2},~\ref{phase2} for
$Re=0.6$ have not yet reached their asymptotic large-$Re$-form
--- cf. also the bifurcation thresholds $r_c^U$ in Fig.~\ref{r_psi}.
For large $Re$ the modulus $|\widehat{c}(z)|$ decreases substantially
and becomes more flat, as was observed \cite{cj93} explicitly for
smaller Soret coupling $\psi=-0.001,~-0.01$.  

\subsubsection{"Stationary" patterns}
Here we discuss the effect of through-flow on stationary patterns at a
moderately positive Soret coupling, $\psi=0.01$, as a representative 
example. In the absence of through-flow, $Re=0$, the perturbation fields
(top part of Fig.~\ref{iso3} and full lines of Fig.~\ref{phase3}) are
real, in phase, and laterally mirror symmetric around the vertical
lines of maximal upflow ($x=0$) and downflow ($x=\pm 1$).
Thus, for $Re=0$ the lateral locations of largest alcohol surplus
(deficiency) coincide with the largest vertical upflow (downflow)
velocity that feeds the bulk with warm, alcohol rich (cold,  alcohol poor)
fluid from the bottom (top) plate.
Note that for $\psi > 0$ the Soret effect causes concentration surplus 
(deficiency) at the warm (cold) plate. The critical convective 
concentration amplitude $|\widehat{c}(z)|$ is for $Re=0$ so large that
the full line representing it in Fig.~\ref{phase3} lies outside the
chosen plot range. The critical wavelength 
$\lambda_c^{(0)}(\psi=0.01) = 2.9833$ of the $Re=0$ pattern
(top part in Fig.~\ref{iso3}) is substantially larger than $2$.

Already a small through-flow changes the above described field 
structure of stationary convective perturbations dramatically.
The wavelength decreases toward $2$. The flow amplitudes
$|\widehat{w}(z)|$ and even more conspicuously the 
concentration amplitude $|\widehat{c}(z)|$ decrease. The phase
$\varphi_c(z)$ of the concentration field exhibits a strong
vertical variation reflecting the almost passive advection by the
flow that is roughly characterized by the parabolic lateral 
through-flow profile superimposed upon a roll-like closed-flow
pattern.

For $\psi<0$, however, amplitudes and phases
of $S$-patterns become more and more similar to those of
$D$-patterns when the two
bifurcation branches $r_c^S$ and $r_c^D$ approach each other with
increasing $Re$ in Fig.~\ref{3d}.
However, the {\it critical} wave numbers of these two structures differ:
For $\psi=-0.001$ we find $\widehat{k}_c^D = 1.001$ and
$\widehat{k}_c^S = 0.998$ and for $\psi=-0.01$ the difference
of the critical wave numbers 
($\widehat{k}_c^D = 1.089,~\widehat{k}_c^S = 0.9678$)
increases. Thus, $\widehat{k}_c^D-\widehat{k}_c^S$
increases with $|\psi|$ when the bifurcation branches collide as described
in Secs.~\ref{iv.c.4} and ~\ref{iv.c.5}.

%%%%%%%%%%%%%%%%%%%%%%%%%%%%%%%%%%%%%%%%%%%%%%%%%%%%%%%%%%%%%%%%%%%%
% \input{v1204ml_2.tex}
%%%%%%%%%%%%%%%%%%%%%%%%%%%%%%%%%%%%%%%%%%%%%%%%%%%%%%%%%%%%%%%%%%%%

\section{ABSOLUTE AND CONVECTIVE INSTABILITY}\label{v}
Whenever at a stability threshold the frequency is nonzero
with a finite group velocity
\begin{equation}
v_g = \left. \frac{\partial \omega(k)}{\partial k} \right|_{k_c}
\end{equation}
one has to distinguish between spatiotemporal growth behavior
of spatially extended and of spatially localized perturbations.
The former having a form $\sim e^{i k x}$ have a positive 
growth rate above the bifurcation thresholds $r_c$ determined 
in the previous section.

\subsection{Wave packets, front propagation, and saddle point analysis}\label{v.a}
A spatially localized perturbation, i.e., a wave packet superposition of
plane wave extended perturbations of a particular kind moves in the so
called convectively unstable parameter regime
\cite{b75,b64,h87,h88} with the velocity $v_g$ 
faster away than it grows --- while growing in the frame comoving with
$v_g$ the packet moves out of the system so that the basic conductive
state is restored. In other words, the two fronts that join the wave packet's
intensity envelope to the structureless state propagate both in the 
direction in which the packet center moves.
On the other hand, in the so called absolutely unstable parameter regime 
the growth rate of the packet is so large that one front propagates 
opposite to the center motion. Thus the packet expands in the 
laboratory frame into the direction of packet motion as well as 
opposite to it \cite{b75,b64}.

We should like to emphasize that we are dealing here only with a
linear analysis of the convective fields. Thus we do not address the 
question whether, e.g., at a subcritical bifurcation the above decribed wave 
packet grows to a stable nonlinear state that then expands back into the 
system with a larger nonlinear front velocity so that the nonlinear
structured state ultimately invades the region occupied by the
homogeneous state.

The boundary in parameter space between convective and 
absolute instability is marked by parameter combinations for which
one of the fronts of the linear wave packet reverts its propagation
direction in the laboratory frame: In the convectively unstable
regime this front propagates in the same direction as the center of the
packet, in the absolutely unstable regime it moves opposite to it, and
right on the boundary between the two regimes the front is stationary
in the laboratory frame. This parameter combination can be determined 
by a saddle point analysis of the linear complex dispersion relation
$s(k)$ over the complex $k$ plane \cite{h87}. Here we do not display the
dependence of $s$ on the control parameters $r,~Re,$ and $\psi$.

The condition of vanishing front propagation velocity is equivalent
to finding the parameters for which
\begin{equation}
\Re s(\kappa) = 0
\label{ca1}
\end{equation}
with $\kappa$ denoting the appropriate saddle position of $s(k)$ determined
by solving
\begin{equation}
\frac{\partial s(\kappa)}{\partial \kappa} = 0
\label{ca2}
\end{equation}
in the complex $k$ plane \cite{h87}.
In Appendix~\ref{appa.2} we describe our numerical method of finding 
the solution of (\ref{ca1},~\ref{ca2}). It yields
the sought after surface in $r-Re-\psi$ parameter space, e.g.,
in the form of a function $r_{c-a}(Re,\psi)$ depending on $Re$ and $\psi$.
This border $r_{c-a}$ lies above the bifurcation threshold
$r_c$ for growth of extended states. Thus for $r<r_c$ the basic conductive
state is stable, for $r_c<r<r_{c-a}$ it is convectively unstable, 
and for  $r_{c-a}<r$ it is absolutely unstable.

\subsection{Ginzburg Landau amplitude equation approximation}\label{v.b}
To solve (\ref{ca1},~\ref{ca2}) for $r_{c-a}$ one has to determine
the dispersion relation $s(k;r,Re,\psi)$ for complex $k$, i.e., one has
to solve the eigenvalue problem (\ref{linpr}) for complex $k$. This is
in general a quite involved numerical task. A somewhat simpler, yet
approximate method, is to use an expansion of $s(k; r, Re, \psi)$
that corresponds to approximate the full field equations by the
Ginzburg Landau amplitude equation (GLE). We recapitulate the 
derivation of the relevant equations here for completeness. The
comparison with results from the full field equation (Sec.~\ref{v.c})
shows that the GLE yields quite useful approximations to the
borderlines between absolute and convective instability. 

\subsubsection{Expansion of $s(k;r)$}
Consider $Re$ and $\psi$ to be fixed for the moment so that we do not have
to display them explicitly in the argument list of $s$. Under the
assumptions {\it (i)} that the sought after saddle $\kappa$ in the 
complex $k$ plane lies close to the critical wave number $k_c$ 
and {\it (ii)} that the relative distance
\begin{equation}
\mu_{c-a} = \frac{r_{c-a}}{r_c} - 1
\end{equation}
between the convective/absolute border $r_{c-a}$ and the critical Rayleigh
number $r_c$ is small we expand 
\begin{eqnarray}
s(k;r) &=&  s_c + ( k - k_c )\, 
\left( \frac{\partial s}{\partial k} \right)_c 
+ \frac{1}{2}\, ( k - k_c )^2 \, 
\left( \frac{\partial^2 s}{\partial k^2} \right)_c \nonumber \\
& & + \mu \, \left( r\, \frac{\partial s}{\partial r} \right)_c + \, 
\mbox{h.~o.~t.} \quad.
\label{entdisp}
\end{eqnarray}
Here we have introduced for convenience the relative distance 
\begin{equation}
\mu = \frac{r}{r_c} - 1
\end{equation}
of the Rayleigh number $r$ from its critical value $r_c$ for onset
of convection. The higher order terms in (\ref{entdisp}) should be of
order $\mu^{3/2}$ since for small $0< \mu \ll 1$ only extended 
perturbations with (real) wave numbers out of a band of width
$k-k_c \sim \sqrt{\mu}$ can grow. 
 
\subsubsection{Relation to linear amplitude equation}
The expansion coefficients of (\ref{entdisp}) appear also in the linear
parts of the complex GLE
\begin{eqnarray}
\tau_0 \left( \partial_t + v_g\, \partial_x \right) \, A  & = &
\left[ \mu\, ( 1 + i c_0 ) + \xi_0^2\, ( 1 + i c_1 )\,
\partial_x^2 \right] \, A \nonumber \\
& & \quad + \quad \mbox{nonlinear terms}.
\end{eqnarray}
Here $A(x,t)$ is the common complex amplitude of convection fields
$\Phi=(w,\theta,c)$
\begin{equation}
\Phi(x,z,t)=A(x,t)\; \widehat{\Phi}(z)\; 
e^{i\, ( k_c x - \omega t )}\; + \; c.c.
\label{ampl_phi}
\end{equation}
that bifurcate out of the conductive state at $\mu=0$. The approximation
(\ref{ampl_phi}) can be expected to be a good one as long as $A$ is small
and, more importantly, as long as the spatial field structure is well
represented by that of the critical eigenfunctions
$\widehat{\Phi}(z)\,e^{i\, k_c x }$.

The relations between the expansion
coefficients of $s$ and the coefficients in the amplitude equation are
\begin{mathletters}
\begin{eqnarray}
\left( \frac{\partial s}{\partial k} \right)_c & = & 
- i\, \left( \frac{\partial \omega}{\partial k} \right)_c   =  - i\, v_g \\
\left(r\, \frac{\partial s}{\partial r} \right)_c & = & 
r_c\, \left(\frac{\partial \gamma}{\partial r} -
i\, \frac{\partial \omega}{\partial r} \right)_c   = 
\frac{1 + i\, c_0}{\tau_0} \\
\label{coeff_c}
\left( \frac{\partial^2 s}{\partial k^2} \right)_c & = &
- \frac{2\, \xi_0^2}{\tau_0}\, ( 1 + i\, c_1 ) \;.
\end{eqnarray}
\label{coeff}
\end{mathletters} 
In (\ref{coeff_c}) we have used the relation \cite{new}
\begin{equation}
\left( \frac{\partial^2 \gamma}{\partial k^2} \right)_c =
- \left( \frac{\partial \gamma}{\partial r} \, 
\frac{\partial^2 r_{stab}}{\partial k^2} \right)_c 
\end{equation}
to relate the second $k$ derivative of the growth rate to the critical
curvature $\xi_0^2 = \frac{1}{2} 
\left(\frac{\partial^2 r_{stab}}{\partial k^2} \right)_c$ of the marginal
stability curve $r_{stab}(k)$. Note that the approximated dispersion
relation (\ref{entdisp}) is precisely the one of the GLE approximation.
 
\subsubsection{Saddle and convective/absolute instability border}
With the notation (\ref{coeff}) the saddle $\kappa$ (\ref{ca2})
of the approximated dispersion (\ref{entdisp}) lies at
\begin{equation}
\kappa = k_c - \frac{\left( \frac{\partial s}{\partial k} \right)_c}
                    {\left( \frac{\partial^2 s}{\partial k^2} \right)_c} =
k_c - \frac{i}{1+i\,c_1} \, \frac{v_g\, \tau_0}{2\, \xi_0^2}\;\;.
\end{equation}
Then the condition (\ref{ca1})
\begin{equation}
0 = \Re\, \left[ \mu \left( r\, \frac{\partial s}{\partial r} \right)_c - 
\frac{1}{2}\, \frac{\left( \frac{\partial s}{\partial k} \right)_c^2}
                   {\left( \frac{\partial^2 s}{\partial k^2} \right)_c}
           \right]
\end{equation}
yields the GLE approximation
\begin{mathletters} 
\begin{equation}
\mu_{c-a} = \frac{v_g^2\, \tau_0^2}{4\, \xi_0^2\, ( 1 + c_1^2 )} \\
\end{equation}
or
\begin{equation}
r_{c-a}    = ( 1 + \mu_{c-a} )\, r_c
\end{equation}
\label{borders}
\end{mathletters} for the boundary between the convectively and absolutely
unstable parameter regime \cite{h87,h88,D85}.

\subsection{Convective and absolute instability against $S,~U,$ and $D$
 perturbations}\label{v.c}

Here we present in Fig.~\ref{konab} our numerical results for the borderlines
between absolute and convective instability obtained from the full
field equations (cf. Appendix~\ref{appa.2}) in comparison with the GLE results. 

There are three different types of extended perturbations 
in our systems, namely $S,~U,$ and $D$, against which the basic
conductive state becomes unstable at the bifurcation threshold 
$r_c^S,~r_c^U$, and $r_c^D$ that have been determined in Sec.~\ref{iv}.
Consequently one has to investigate the spatiotemporal growth behavior
of three different packets consisting of superpositions of $S,~U,$ or $D$
plane wave perturbations. Their analysis along the lines of Sec.~\ref{v.a} 
and Appendix~\ref{appa.2} yields
three functions $r_{c-a}^S$ (circles), $r_{c-a}^U$ (upwards pointing 
triangles),
and $r_{c-a}^D$ (downwards pointing triangles) each depending on
$Re$ and $\psi$ that mark the boundary surfaces in $r-Re-\psi$ parameter 
space between the convectively and absolutely regime of the basic state against
type $S,~U,$ or $D$ perturbations.

In order to determine the three boundaries within the GLE approximation 
we have determined the derivatives (\ref{coeff}) of the respective eigenvalues 
$s^S,~s^U,$ and $s^D$ at their respective critical points
$r_c(Re,\psi),~k_c(Re,\psi)$ \cite{cj93}.
Then (\ref{borders}) yields the functions
$r_{c-a}(Re,\psi)$ for the three patterns.
They are shown in Fig.~\ref{konab} for a few $\psi$
as functions of $Re$ by thick lines 
($S$: dotted,~$U$: full,~$D$: dashed) together with the corresponding bifurcation 
thresholds $r_c$ (thin lines).
For positive Soret coupling $\psi=0.01$, $0.001$ one obtains
within the GLE approximation a local maximum
of $r^S_{c-a}$ ($\psi=0.01$ at $Re\approx 0.25$, $\psi=0.001$ at $Re\approx 0.1$),
when the product of $\tau_0$ decreasing with $Re$ and $v_g$ increasing with $Re$
reaches a maximum.
In the $\psi=0.01$ case a second maximum is visible at small $Re$ 
that is possibly associated with a root of $c_1$. 
In contrast $r^S_{c-a}$ of the full field equations (circles) is
monotonously increasing with $Re$.
For negative Soret coupling $\psi=-0.001$, $-0.01$, $-0.1$, and $-0.25$ 
$r^U_{c-a}$ first decreases with increasing $Re$, coincides at its local
 minimum with $r^U_c$
when $v_g=0$, and afterwards increases with $Re$. The GLE approximation 
yields good quantitative agreement with the results of the full field equations
for $r^U_{c-a}$ (upwards pointing triangles)
in the vicinity of the local minimum, while for higher $Re$ the GLE results
increase more strongly.
For smaller Reynolds numbers this validity range is typically enlarged down to
$Re=0$ for the $\psi$-values presented here.

As discussed in Sec.~\ref{iv.c.5} we have limited our investigation
of $r^D_c$ and $r^S_c$ as functions of $Re$ to cases, where the wave number gap
in the $D-S$ marginal stability curves has
not yet appeared. Therefore, within the GLE approximation $r^D_{c-a}$ and
 $r^S_{c-a}$
are determined by the expansion coefficients of $s$ (\ref{coeff}) at the 
critical Rayleigh number $r_c$
only as long as the calculation of $r^D_c$ and $r^S_c$ was numerically possible. 
Nevertheless, for higher $Re$ critical Rayleigh numbers still exist to the left
and right hand side of the wave number gap (cf. Fig.\ref{OM}). 
We also found that the $S$ saddle point of the full
field equations (evaluated by
succesively increasing $Re$) evolves
monotonously when $Re$ increases beyond the threshold where  
the wave number gap occurs.
The wave number of this saddle point is always above $\hat{k}=1$ and increases
with $Re$.

The GLE results for $r^D_{c-a}$ are increasing stronger than those
of the full field equations (downwards pointing triangles).
For $\psi=-0.001$ the $r^D_{c-a}$ stability
limit seems to terminate close to the Reynolds number
where the wave number gap (cf. Sec.~\ref{iv.c.5}) opens up in
the $D-S$ marginal stability curves. 
There $\Im\kappa$ seems to change sign when 
increasing $Re$ further. A graphical analysis of this saddle that has moved
into the lower complex $k$-plane suggests that there
$\Re s < 0$ above a certain $Re$-limit for all $r$ so that this saddle can
be ignored.

%%%%%%%%%%%%%%%%%%%%%%%%%%%%%%%%%%%%%%%%%%%%%%%%%%%%%%%%%%%%%%%%%%%%
% \input{vi1104ml.tex}
%%%%%%%%%%%%%%%%%%%%%%%%%%%%%%%%%%%%%%%%%%%%%%%%%%%%%%%%%%%%%%%%%%%%

\section{conclusion}
We have investigated the influence of an externally imposed horizontal
shear flow on linear convective structure formation in binary fluid
layers heated from below. To that end we have solved the linearized 
field equations for convective perturbations of the basic conductive 
state numerically with a shooting method. In addition we have checked 
our results --- in particular some of the peculiar changes wih $Re$ ---
by a variational calculation that gave a good agreement. 
We have determined for positive and negative Soret coupling $\psi$
the $Re$-dependence of the critical bifurcation properties: stability
thresholds, wave numbers, frequencies, and eigenfunctions for
three different types of perturbations. The latter are identified 
by different characteristic exponents that cause perturbations
to be stationary ($S$), downstream traveling ($D$), or upstream
traveling ($U$) at $Re=0$. 

The Hopf symmetry degeneracy of $U$ and $D$ perturbations
at $Re=0$ and $\psi<0$ is broken by a finite through-flow --- wave numbers
$k_c^U$ and $k_c^D$, frequencies $\omega_c^U$ and $\omega_c^D$, and
bifurcation thresholds $r_c^U$ and $r_c^D$ develop differences. At
small $Re$ upstream traveling wave convection grows first upon
increasing the heating since $r_c^U$ is depressed and $r_c^D$
is shifted upwards. But then, with increasing through-flow the 
bifurcation lines $r_c^U$ and $r_c^D$ intersect giving rise to
bistable bifurcation behavior. Eventually $r_c^U$ flattens out and
approaches for large $Re$ the pure fluid's stability boundary 
$r_c^S(Re,\psi=0)$ --- a sufficiently strong shear flow eliminates
the Soret induced coupling effects between the convective
concentration field and temperature and velocity fields. This
also holds for mixtures with positive $\psi$. The bifurcation
thresholds $r_c^D$ monotonously curve upwards when increasing $Re$
and collide with the $r_c ^S$ threshold lines
that for $\psi<0$ sharply drop downwards with growing $Re$.
This behavior is easily understood by analyzing the $Re$-variation
of the paths of the relevant three eigenvalues in the complex
$\gamma - \omega$ plane and of the variation of $\gamma(k,r)$.

At $\psi>0$ the stationary threshold $r_c^S$ rapidly approaches the
$\psi=0$ asymptote of a pure fluid from below --- again the shear
flow suppresses
the vertical convective transport of Soret induced concentration 
gradients. Thus, any Soret effect's influence on the bifurcation
thresholds is eliminated by sufficiently large through-flow.

Finally we have evaluated the borders  $r_{c-a}^U$, $r_{c-a}^D$, and
$r_{c-a}^S$ between convective and absolute instability for $U,~D$, and
$S$ perturbations, respectively. To that end we have determined the 
relevant saddles
of the dispersion relations $s^U(k)$, $s^D(k)$, and $s^S(k)$ in
the complex wave number plane. These numerically exact results were
compared with GLE approximations which agree reasonably well with the
former. The latter were obtained by an 
expansion around the respective critical values and by evaluating 
all the coefficients that enter into the linear GLE for $U,~D$, and $S$
patterns.

\acknowledgments
Support by the Deutsche Forschungsgemeinschaft is gratefully acknowledged.

%%%%%%%%%%%%%%%%%%%%%%%%%%%%%%%%%%%%%%%%%%%%%%%%%%%%%%%%%%%%%%%%%%%%
% \input{appa1204pb.tex}
%%%%%%%%%%%%%%%%%%%%%%%%%%%%%%%%%%%%%%%%%%%%%%%%%%%%%%%%%%%%%%%%%%%%

\appendix
\section{Numerical analysis}\label{appa}
\subsection{Shooting algorithm}\label{appa.1}
Here we describe our version of the shooting algorithm modified such
as to better cope with the numerical problems caused by the
concentration boundary layers near the horizontal plates.

The equations (\ref{alleqs}) are written as an $8$-D
system of first order differential equations
\begin{equation}
\partial_z{\bf y} - {\cal A}(z,\mbox{{\bf $\lambda$}})\,
{\bf y} = 0 \quad \mbox{at} \quad 0 \le z \le 1 \quad \quad.
\label{a1}
\end{equation}
where the coefficient matrix follows directly from (\ref{linpr}).
The eigenvalues 
\begin{equation}
\mbox{{\bf $\lambda$}} = 
\left[ r_{stab}(k),~\omega(k) \right]_{\gamma=0}
\quad \mbox{or} \quad 
\mbox{{\bf $\lambda$}} = 
\left[ \gamma(k),~\omega(k) \right]_{r\; \mbox{{\tiny{fixed}}}}
\end{equation}
are determined as a function of $k$ either for $\gamma=0$ on the 
marginal stability boundary $r_{stab}(k)$ or for fixed $r$.
The eigenfunctions
\begin{equation}
{\bf y} = ( w,~\partial_z w,~\partial_z^2 w,~\partial_z^3 
w,~\theta,~\partial_z \theta,~\zeta,~\partial_z \zeta )
\end{equation}
fulfill the boundary conditions 
\begin{equation}
y_1 = y_2 = y_5 = y_8 = 0 \quad \mbox{at} \quad z = 0,\,1 \quad.
\label{bc_b}
\end{equation}
following from (\ref{randbed}).

First (\ref{a1}) is integrated with a fourth-order Runge-Kutta method with the
initial vectors 
\begin{equation}
{\bf y}^1 = \hat{{\bf e}}_3,\;
{\bf y}^2 = \hat{{\bf e}}_4,\;
{\bf y}^3 = \hat{{\bf e}}_6,\; \mbox{and}\;
{\bf y}^4 = \hat{{\bf e}}_7
\end{equation}
where $\hat{{\bf e}}_j$ is the unit vector in $j$-direction.
Then an arbitrary linear combination 
\begin{equation}
{\bf y} = \sum \limits_{j=1}^4 \, a_j\, {\bf y}^j
\end{equation}
fulfills the boundary conditions at $z=0$, but not at $z=1$. The latter
is guaranteed by choosing the correct coefficients $a_j$. This leads to a set
of homogeneous equations
\begin{equation}
\left( \left.
\begin{array}{cccc}
y_1^1 & y_1^2 & y_1^3 & y_1^4 \\
y_2^1 & y_2^2 & y_2^3 & y_2^4 \\
y_5^1 & y_5^2 & y_5^3 & y_5^4 \\
y_8^1 & y_8^2 & y_8^3 & y_8^4
\end{array} \right) \right| _{z = 1}
\; \cdot \left( \begin{array}{c}
a_1 \\ a_2 \\ a_3 \\ a_4
\end{array} \right) \; = \; {\bf 0}  \quad.
\label{a7}
\end{equation}
The nontrivial solution of (\ref{a7}) requires the determinant of the
coefficients to vanish. Thus, this solvability condition but also the
explicit evaluation of the eigenvalues requires the knowledge of the 
eigenfunctions at $z=1$. This is a problem
because of the strong variations in the boundary 
layers --- see, e.g., Fig.~\ref{iso1}.
Our integrator is not able to shoot exactly from $z=0$ to $z=1$ so that
we can not evaluate the eigenvalues $\lambda$ with 
a fixed accuracy.
Especially, this fact makes it 
difficult to find a good critical wave number. To understand this,
we can expand the stability curve $r_{stab}$ around $k_c^{exact}$
\begin{eqnarray}
\Delta r &=& r_{stab}( k_c^{exact} + q )-r_{stab}(k_c^{exact}) \nonumber \\
    &=& \frac{1}{2}\, q^2\, \left. \frac{\partial^2 r_{stab}}{\partial k^2}
\right|_{k_c^{exact}} + O(q^3)\quad
\end{eqnarray}
and we find the defect
\begin{equation}
q \approx \sqrt{ \frac{\Delta r}{\xi_0^2} } \quad.
\end{equation}
With $\xi_0^2 \sim O(1)$ the accuracy $\Delta r$ strongly affects
the accuracy of the critical wave number and of the coefficients
entering the GLE.

One could change the integrator. But the problem can be solved more easily.
Using the mirror symmetry of the eigenfunctions ($w,~\theta,~\zeta$)
one only needs to shoot from $z=0$ to mid height $z=1/2$. There the new
boundary conditions are
\begin{equation}
y_2 = y_4 = y_6 = y_8 = 0 \quad \mbox{at} \quad z = \frac{1}{2} \quad.
\end{equation}
The new homogeneous equations replacing (\ref{a7}) are then 
\begin{equation}
\left( \left.
\begin{array}{cccc}
y_2^1 & y_2^2 & y_2^3 & y_2^4 \\
y_4^1 & y_4^2 & y_4^3 & y_4^4 \\
y_6^1 & y_6^2 & y_6^3 & y_6^4 \\
y_8^1 & y_8^2 & y_8^3 & y_8^4
\end{array} \right) \right| _{z=\frac{1}{2}}
\; \cdot \left( \begin{array}{c}
a_1 \\ a_2 \\ a_3 \\ a_4
\end{array} \right) \; = \; {\bf 0} \quad.
\end{equation}
Now, shooting only from $z=0$ to $z=1/2$ we can compute the eigenvalues 
with a fixed accuracy in our parameter range. 
A welcome secondary effect is that one saves CPU-time.
However, it must be said that this trick only delays the boundary-layer-problem
without eliminating it completely:
If the parameters are such that the phase velocity $|\omega_c/k_c|$
reaches about $20$ the boundary-layer at $z=0$ has become so thin,
that it is no more possible to use our integrator. 

\subsection{Saddle point analysis}\label{appa.2}
The borderline $r_{c-a}$ between convective and absolute instability is
defined by the conditions (\ref{ca1},\ref{ca2}) 
on the dispersion relation after a 
continuation into the complex wave number plane, which also
gives the saddle point position $\kappa$ and the frequency $\omega(\kappa)$.
We have determined $\Re s(k)$ and $\Im s(k)$ with a shooting method
from $z=0$ to mid height $z=1/2$ (see Appendix~\ref{appa.1}) for given 
$r$, $Re$, and $\psi$ and {\em complex} wave number $k$.
Then we solved for fixed $Re$ and $\psi$ 
the nonlinear system of the three equations
\begin{equation}
\Re s(\kappa,r) = 0,
\end{equation}
\begin{equation}
\frac{\partial \Re s(\kappa,r)}{\partial (\Re \kappa)} = 0
\end{equation}
\begin{equation}
\frac{\partial \Re s(\kappa,r)}{\partial (\Im \kappa)} = 0
\end{equation}
that correspond to (\ref{ca1},\ref{ca2}) with 
a Newton-Raphson method with backtracking. The solution yields the borderline 
value $r_{c-a}$ and the saddle position $\kappa$ of $s$ in the 
complex $k$-plane.

The Jacobian matrix as well as the partial derivatives were obtained in  
discretized form by using central differences in the variables $\Re k$,
$\Im k$, and $r$. Mainly two problems occur: one needs a good initial guess
to ensure convergence of the Newton-Raphson method, and for higher 
Reynolds numbers, which are not discussed here, the shooting method
starts to fail for the given accuracy limit (see Appendix~\ref{appa.1}).

We also used an alternative iterative method to evaluate the saddle points.
It yields the same results as our first method, but
it does not require partial derivatives.
First we calculate the eigenvalue 
\begin{equation}
\left[\; \omega(\widetilde{\kappa}),\, r(\widetilde{\kappa})\; \right]_{\gamma=0}
\end{equation}
at a suitably chosen initial value $\widetilde{\kappa}$ for the saddle position.
Then, along a circle in the complex $k$-plane of radius $\rho$ around
$\widetilde{\kappa}$ 
we determine, say, $n \simeq 10$ values 
\begin{equation}
\left[\; \omega(k_j),\, \gamma(k_j)\; \right]_{r=r(\widetilde{\kappa})}
\quad,\quad j=0,..,n-1
\end{equation}
with
\begin{equation}
k_j=\widetilde{\kappa} + \rho\, \left[\; \cos \left( \frac{2\,j\,\pi}{n} \right) 
+ i\; \sin \left( \frac{2\,j\,\pi}{n} \right) \; \right]\;.
\end{equation}
The function values $\{\omega(\widetilde{\kappa}),\omega(k_0),..,\omega(k_{n-1})\}$
are then used for a biquadratical fit
\begin{eqnarray}
\omega(k) &=& a_0 + a_1\,\Im k + a_2\,\Re k + a_3\, (\Im k)^2 \nonumber \\ 
& & + a_4\, (\Re k)^2 + a_5\,\Im k\,\Re k
\label{biqu}
\end{eqnarray}
and the analytically determined saddle of (\ref{biqu}) is used 
as a new initial value $\widetilde{\kappa}$. Now, we reduce the circle radius
$\rho$ by a factor $\alpha$ and continue as above. We repeat this procedure
until the changes of $\widetilde{\kappa}$ have fallen below a suitable limit.
This method is fast, robust, and easily programmable. 
However, the first $\rho$ and the 
reducing factor $\alpha$ have to be chosen carefully. Typically we put
$\rho \sim O(10^{-2})$ and $\alpha=\sqrt 2$.

%%%%%%%%%%%%%%%%%%%%%%%%%%%%%%%%%%%%%%%%%%%%%%%%%%%%%%%%%%%%%%%%%%%%
% \input{appb2803ml.tex}
%%%%%%%%%%%%%%%%%%%%%%%%%%%%%%%%%%%%%%%%%%%%%%%%%%%%%%%%%%%%%%%%%%%%

\section{Comparison with a variational method}\label{appb}
In view of the unexpected $Re$ dependence, say, of Fig.~\ref{krw_01} 
we wanted to compare our results obtained with a shooting method
with a completely different and independent method. To that end
we used the variational method of Prigogine and Glandsdorff 
\cite{gp64,p71-1}. While this method does not yield
exact results it still is very valuable in providing an independent
check for our numerical analysis.

The aim of this variational method is to find an functional, called
local potential, and to minimize it.
We start with the linearized field-equations
\begin{mathletters}
\label{apba}
\begin{eqnarray}
\partial_tu&=&-\sigma\, Re\, P\, \partial_xu 
-\partial_x p + \sigma \nabla^2 u \nonumber \\
 & &\quad -\sigma\, Re\, w \partial_z P \label{a11}  \\
\partial_tw&=&-\sigma\, Re\, P\, \partial_xw -
\partial_zp + \sigma \nabla^2 w \nonumber \\
 & &\quad + \sigma \left[ ( 1+\psi) \theta + \zeta\right] \label{a2} \\
\partial_t\theta &=& -\sigma\, Re\, P\, \partial_x \theta +
\nabla^2 \theta + Ra\,w 
\label{a3} \\
\partial_t\zeta&=&-\sigma\, Re\, P\, \partial_x \zeta +
L \nabla^2 \zeta 
- \psi \nabla^2 \theta  
\label{a4} \\
0&=& \partial_x u + \partial_z w \label{a5} \quad.
\end{eqnarray}
\end{mathletters}
Here $u,\, w,\, \theta,\, \zeta $ and $p$ are convective disturbances from the 
conductive profiles and $P(z)$ the Poiseuille shear flow (\ref{ucondb}).
Following the procedure explained in Ref.~\cite{gp64} we multiply the above
equations by increments of the respective fields:
(\ref{a11}) by $-\delta u$, (\ref{a2}) by $-\delta w$,
(\ref{a3}) by $-\delta \theta$ and (\ref{a4}) by $-\delta\zeta$.
Here, e.g.,
\begin{equation}
w = w^0 + \delta w
\end{equation}
with $w^0$ being a solution of (\ref{apba}) that is not varied in the following.
Then we add and obtain
\begin{eqnarray}
& &-\frac{1}{2} \partial_t \left[ (\delta u)^2 + (\delta w)^2 +
 (\delta \theta)^2
+ (\delta \zeta)^2 \right] \nonumber \\
&=& \partial_t u^0\, \delta u + \partial_t w^0\, \delta w +
    \partial_t \theta^0 \, \delta \theta + \partial_t \zeta^0 \,
    \delta \zeta \nonumber \\
& & + \sigma\,Re\,P\, \partial_xu \, \delta u +
    \sigma\, Re \,
w\, \partial_zP\, \delta u \nonumber \\
& & + \partial_xp \, \delta u
-\sigma\, \delta u\, \nabla^2 u 
+ \sigma\,Re\,P\, \partial_xw \, \delta w \nonumber \\
& & - \sigma\, \delta w \left[(1+\psi)\theta + \zeta\right]
-\sigma\, \delta w \nabla^2 w 
- Ra\, w\, \delta\theta \nonumber \\
& &  - \delta \theta\, \nabla^2 \theta +
\sigma\, Re \,P\, \partial_x\theta \, 
\delta \theta + \partial_zp\,\delta w \nonumber \\
& & - L\,\delta \zeta\,  \nabla^2 \zeta
+ \psi \, \delta \zeta\, \nabla^2 \theta
+ \sigma\, Re\, P\, \partial_x \zeta \, \delta \zeta 
\end{eqnarray}
when making use of relations like
\begin{equation}
-\partial_t w \,\delta w = -\partial_t \left(w^0+\delta w \right)\, 
\delta w = -\frac{1}{2} \partial_t (\delta w)^2 -
 \partial_t w^0\, \delta w \quad.
\end{equation}
We expand the fields in plane waves, $\Phi=\widehat{\Phi}(z) e^{i k x}e^{s t}$,
and we integrate over the entire $x-z$ crosssection of the fluid layer.
Following closely the prescriptions of Ref.~\cite{p71-1} we then
otain the local potential 
\begin{eqnarray}
\Psi &=& \int\limits_0^1 \, dz \, 
\left[\;\; \frac{s}{k^2}\, \partial_z\widehat{w}^0\, 
\partial_z\widehat{w} \right. + s\; \left( \widehat{w}^0\,\widehat{w} +
\widehat{\theta}^0\, \widehat{\theta} + \widehat{\zeta}^0\, 
\widehat{\zeta} \right) \nonumber \\
& & + \sigma\, \partial_z\widehat{w}^0\, \partial_z\widehat{w} 
- \frac{\sigma}{2\, k^2}\, \left( \partial_z^2\widehat{w} \right)^2 
+\frac{2\, \sigma}{k^2}\, \partial_z^2\widehat{w}^0\, \partial_z^2\widehat{w} 
\nonumber \\
& &  + i\, k\, \sigma\,Re\, P \left( \widehat{w}^0\, \widehat{w} +
\widehat{\theta}^0\, \widehat{\theta} + \widehat{\zeta}^0\,
\widehat{\zeta} \right) 
+ \frac{\sigma}{2}\, \left( \partial_z\widehat{w} \right)^2  \nonumber \\ 
& & + \sigma\, k^2 \left( 
\widehat{w}^0\, \widehat{w} + \frac{1}{\sigma} \widehat{\theta}^0\,
\widehat{\theta} + \frac{L}{\sigma} \widehat{\zeta}^0\,
\widehat{\zeta} \right)
-\sigma (1+\psi) \widehat{\theta}^0\, \widehat{w} \nonumber \\
& & -Ra\, \widehat{w}^0\, \widehat{\theta} +
\frac{1}{2} \left( \partial_z \widehat{\theta} 
\right)^2 + \frac{L}{2} \left( \partial_z \widehat{\zeta} \right)^2
- \sigma\, \widehat{\zeta}^0\, \widehat{w}
\nonumber \\
& & - \psi \, k^2\,
\widehat{\theta}^0\, \widehat{\zeta} 
+ \psi\, \partial_z^2\widehat{\theta}^0\, \widehat{\zeta} 
- \frac{i}{k} \, \sigma\, Re\, \widehat{w}^0\,
\partial_z\widehat{w}\, \partial_zP \nonumber \\
& & + \left. \frac{i}{k}\, \sigma\, Re\, P\, \partial_z\widehat{w}^0\, 
\partial_z\widehat{w}\;\; \right]\,. 
\end{eqnarray}
Here we have elimitated the pressure by using (\ref{a11}) 
\begin{eqnarray}
-\widehat{p}^0 & = &\frac{s}{k^2}\, \partial_z\widehat{w}^0 + \frac{i}{k}\,
 \sigma\, Re\, P\, \partial_z\widehat{w}^0 \nonumber \\
& &\; - \frac{i}{k}\, \sigma\, Re\, \widehat{w}^0\,
\partial_zP - \frac{\sigma}{k^2}\, \partial_z^3\widehat{w}^0 +
\sigma\, \partial_z\widehat{w}^0 \quad.
\end{eqnarray}
Furthermore, partial integrations in $z$ have been applied.
One can check directly that the Euler-Langrange equation
for, say,  $\widehat{\zeta}$
\begin{equation}
\frac{\partial \Psi}{\partial \widehat{\zeta}} -
\frac{\partial}{\partial z}\, \frac{\partial \Psi}{\partial 
\left( \frac{\partial \widehat{\zeta}}{\partial z} \right)} = 0 
\end{equation}
leads to 
\begin{equation}
\left[ s + i k \sigma Re P \right] \widehat{\zeta}^0 = 
L \left( \partial_z^2\widehat{\zeta} - k^2 \widehat{\zeta}^0 \right) -
\psi \left( \partial_z^2 - k^2 \right) \widehat{\theta}^0 \quad.
\end{equation}
This is together with the a-posteriori subsidiary condition
$\widehat{\zeta} = \widehat{\zeta}^0$ the equation
(\ref{a4}). An analogous calculation leads with respective subsidiary 
conditions $\widehat{\theta} = \widehat{\theta}^0\;,\;
\widehat{w} = \widehat{w}^0$ to the other 
differential equations.

To get the critical values we expand the unknows $\widehat{w},~ 
\widehat{\theta}$ and $~\widehat{\zeta}$ 
\begin{mathletters}
\label{trial1}
\begin{eqnarray}
\widehat{w}  (z) =
 \sum_{j=1}^{N_1} \, \widetilde{w}_j   \, f_j(z) & \quad & 
\widehat{w}^0(z) =
 \sum_{j=1}^{N_1} \, \widetilde{w}_j^0 \, f_j(z) \\
\widehat{\theta}(z) =
 \sum_{j=1}^{N_2} \, \widetilde{\theta}_j \, g_j(z) & \quad & 
\widehat{\theta}^0(z) =
 \sum_{j=1}^{N_2} \, \widetilde{\theta}_j^0 \, g_j(z) \\
\widehat{\zeta}(z) =
 \sum_{j=1}^{N_3} \, \widetilde{\zeta}_j \, h_j(z) & \quad &
\widehat{\zeta}^0(z) =
 \sum_{j=1}^{N_3} \, \widetilde{\zeta}_j^0 \, h_j(z)
\end{eqnarray}
\end{mathletters}
in functions $f_j,~g_j$, and $h_j$ that satisfy the NSI boundary conditions. 
As trial functions we use
\begin{mathletters}
\label{trial2}
\begin{eqnarray}
f_j(z) & = & \left[ z ( 1-z) \right]^{2\,j} \\
g_j(z) & = & z ( 1-z) \, (2\,z -1 )^{2(j-1)} \\
h_j(z) & = & \left[ z(1-z) \right]^{2(j-1)} \quad.
\end{eqnarray}
\end{mathletters}
We insert (\ref{trial1},~\ref{trial2}) into the expression for $\Psi$,
we minimize with respect to the
variational parameters $\widetilde{w}_j, \widetilde{\theta}_j$
and $\widetilde{\zeta}_j$, and we use the subsidiary conditions
\begin{equation}
\widetilde{w}_j = \widetilde{w}_j^0 \;,\;
\widetilde{\theta}_j = \widetilde{\theta}_j^0 \;,\;
\widetilde{\zeta}_j = \widetilde{\zeta}_j^0 \quad.
\end{equation}
Then we obtain a system of $N=N_1+N_2+N_3$ linear homogeneous 
equations ${\cal A}{\bf x}={\bf 0}$ where the vector ${\bf x}$ 
has $N$ components containing $\widetilde{w}_j^0,~\widetilde{\theta}_j^0$,
and $\widetilde{\zeta}_j^0$. For nontrivial solutions to exist 
the determinant of ${\cal A}$ must be zero, so that we have to 
solve the equation
\begin{equation}
\det\, {\cal A} [ r_{stab}(k), \omega (k) ]_{\gamma=0} =
 0 \label{equat}\; .
\end{equation}
With $N_1=1,~N_2=3,~N_3=4$ the software package 
MATHEMATICA is able to solve (\ref{equat}) 
exactly.

The variational results display the same structural
properties as the shooting results and  
they agree within less than $1\%-2\%$ for $Re \neq 0$. For $Re = 0$
the agreement is better. We found, e.g., for $\psi=-0.25$
the critical values $r_c=1.33497, ~\widehat{k}_c=1.00451$,
and $\omega_c=\pm 11.20027$ differing by less than
 $ 0.2\%$ from the shooting results. 
As an interesting aside we mention that the variational
calculus yields in the absence of through-flow 
$\psi_{\infty}^0=-L/(1+L)$ exactly. 
Furthermore, the separation ratio $\psi_0$ for which
$k_c=0$ is obtained as $\psi_0 = L/(f-L)$
with $f=6695603/25739142=0.2601...$ which has to be 
compared with the analytically exact result 
$f=34/131=0.2595...$ \cite{k88,sh95}.

%%%%%%%%%%%%%%%%%%%%%%%%%%%%%%%%%%%%%%%%%%%%%%%%%%%%%%%%%%%%%%%%%%%%
% \input{lit1204cj.tex}
%%%%%%%%%%%%%%%%%%%%%%%%%%%%%%%%%%%%%%%%%%%%%%%%%%%%%%%%%%%%%%%%%%%%

%\end{references}

\newpage
\narrowtext

%%%%%%%%%%%%%%%%%%%%%%%%%%%%%%%%%%%%%%%%%%%%%%%%%%%%%%%%%%%%%%%%%%%%
% \input{fig1204cj.tex}
%%%%%%%%%%%%%%%%%%%%%%%%%%%%%%%%%%%%%%%%%%%%%%%%%%%%%%%%%%%%%%%%%%%%

\addcontentsline{toc}{section}{FIGURES}
%\hspace{4cm} {\large Figures}
%\\ \\
\begin{figure}\caption[]
{Eigenvalues for small Reynolds numbers. The variation of growth rates
$\gamma$ and frequencies $\omega$ with $Re$ resulting from an 
expansion up to linear order in $Re$ is shown schematically. The  
lifting of the symmetry degeneracy, $s_2^{(0)}=s_1^{(0)\,*}$, by
the through-flow and the behavior of $s_3^{(0)}$ is described in
Sec.~\ref{iii.d}.}
\label{skizze1}
\end{figure}
% ============================================================================
\begin{figure}\caption[]
{Evolution of Hopf bifurcation properties with through-flow. Shown are
critical Rayleigh numbers (a), reduced wave numbers (b), and frequencies
(c) for upstream (full lines) and downstream (dashed lines) traveling 
perturbations that are 
symmetry degenerate for $Re=0$. Parameters are $L=0.01,~\sigma=10,~\psi=-0.1$.}
\label{krw_01} 
\end{figure}
% ============================================================================
\begin{figure}\caption[]
{$Re$-dependence of bifurcation thresholds for negative Soret coupling $\psi$.
The stability boundaries $r_c^S$ (dotted lines), $r_c^D$ (dashed lines), and
$r_c^U$ (full lines) are shown for some representative $\psi$-values as 
indicated. The behavior beyond the $Re$ values where $r_c^D$ and $r_c^S$
collide is discussed in Sec.~\ref{iv.c.5}.
Parameters are $L=0.01,~\sigma=10$.}
\label{r_psi} 
\end{figure}
% ============================================================================
\begin{figure}\caption[]
{(a) Frequencies $\omega(r)$ and (b) growth rates $\gamma(r)$ of the two 
eigenvalues that cause collision of the stability curves 
$r_c^D$ and $r_c^S$ in Fig.~\ref{r_psi}. Thick (thin) lines 
and symbols referring to $Re=0$ ($Re=0.4$) were obtained for a fixed
wave number $\widehat{k}=1$. Arrows indicate deformation directions
of the curves. The $Re$-variation of the zeros of $\gamma$ at
$r^U$ (upwards pointing triangle), $r^D$ (downwards pointing triangle),
and $r^S$ (circle) is discussed in Sec.~\ref{iv.c.4}.
Parameters are $L=0.01,~\sigma=10,~\psi=-0.01$.} 
\label{eigw}
\end{figure}
% ============================================================================
\begin{figure}\caption[]
{{\it Schematic} variation of the eigenvalues of Fig.~\ref{eigw} in the complex
$\gamma-\omega$--plane. Arrows indicate the motion of the 
eigenvalues with increasing $r$.
Line styles and symbols are those of Fig.~\ref{eigw}. }
\label{eigw2}
\end{figure}
% ============================================================================
\begin{figure}\caption[]
{The "mountain landscape" of $\gamma^{D/S}$ over the $\widehat{k}-r$ plane for
$Re=0.4$ (a) and $Re=0.475$ (b). Gray-scales show the height of 
$\gamma^{D/S}$ in the range where $\gamma>0$. Thick lines are marginal
stability curves where $\gamma=0$. In the white parts of the figures 
$\gamma^{D/S}<0$. Dashed lines are iso-$\gamma$ lines. The zeros
of $\gamma^{D/S}$ marked by circle and downwards pointing triangle
are those of Fig.~\ref{eigw}b.
Parameters are $L=0.01,~\sigma=10,$ and $\psi=-0.01$.}   
\label{OM}
\end{figure}
% ============================================================================
\begin{figure}\caption[]
{(a) Critical Rayleigh numbers and (b) wave numbers for positive $\psi$
versus through-flow Reynolds numbers. Parameters are $L=0.01,~\sigma=10$.}
\label{rwkpsi0}
\end{figure}
% ============================================================================
\begin{figure}\caption[]
{To give an impression of the bifurcation surfaces in $r-Re-\psi$ space we
show  $r_c^U$ (thin full lines), $r_c^D$ (thin dashed lines), and
$r_c^S$ (thin dotted lines) together with their  $\psi$-dependence for $Re=0$
(thick lines). Parameters are $L=0.01,~\sigma=10$.}
\label{3d}
\end{figure}
% ============================================================================
\begin{figure}\caption[]
{Spatial structure of $D$ patterns that propagate downstream, i.e., 
to the right.
Contours of the fields $w,~\theta,~c$ in vertical cross sections
of the fluid layer are shown for different $Re$ as indicated.
White (black) implies large (small) field values. The largest vertical
up (down) flow is at $x=0 (\pm 1)$. Contour lines mark fractions
$\pm (0,1,2,3,4)/5$ of the field maxima. Parameters are
$L=0.01,~\sigma=10,~\psi=-0.25$. The critical frequencies are
about $\omega_c^D \approx 11.21 + 41.9\,Re$.}
\label{iso1}
\end{figure}
% ============================================================================
\begin{figure}\caption[]
{Vertical profiles of $D$ patterns. 
Shown are moduli (left column) and phases (right column)
of the critical complex amplitudes that are symmetric
around the mid position $z=1/2$. Their normalization 
is explained in Sec.~\ref{iv.f}. Parameters are $L=0.01,~\sigma=10,~\psi=-0.25$.}
\label{phase1}
\end{figure}
% ============================================================================
\begin{figure}\caption[]
{Spatial structure of $U$ patterns. They propagate for $Re \rightarrow 0$
upstream, i.e., to the left.
Contours of the fields $w,~\theta,~c$ in vertical cross sections
of the fluid layer are shown for different $Re$ as indicated.
White (black) implies large (small) field values. The largest vertical
up (down) flow is at $x=0 (\pm 1)$. Contour lines mark fractions
$\pm (0,1,2,3,4)/5$ of the field maxima. Parameters are
$L=0.01,~\sigma=10,~\psi=-0.25$. The critical frequencies are
about $\omega_c^U \approx -11.21 + 41.9\,Re$.}
\label{iso2}
\end{figure}
% ============================================================================
\begin{figure}\caption[]
{Vertical profiles of $U$ patterns. 
Shown are moduli (left column) and phases (right column)
of the critical complex amplitudes that are symmetric
around the mid position $z=1/2$. Their normalization 
is explained in Sec.~\ref{iv.f}. Parameters are $L=0.01,~\sigma=10,~\psi=-0.25$.}
\label{phase2}
\end{figure}
% ============================================================================
\begin{figure}\caption[]
{Spatial structure of $S$ patterns. They are stationary for $Re \rightarrow 0$.
Contours of the fields $w,~\theta,~c$ in vertical cross sections
of the fluid layer are shown for different $Re$ as indicated.
White (black) implies large (small) field values. The largest vertical
up (down) flow is at $x=0 (\pm 1)$. Contour lines mark fractions
$\pm (0,1,2,3,4)/5$ of the field maxima. Parameters are
$L=0.01,~\sigma=10,~\psi=0.01$. The critical frequencies are
about $\omega_c^S \approx 41.9\,Re$.}
\label{iso3}
\end{figure}
% ============================================================================
\begin{figure}\caption[]
{Vertical profiles of $S$ patterns. 
Shown are moduli (left column) and phases (right column)
of the critical complex amplitudes that are symmetric
around the mid position $z=1/2$. Their normalization 
is explained in Sec.~\ref{iv.f}. Parameters are $L=0.01,~\sigma=10,~\psi=0.01$.}
\label{phase3}
\end{figure}
% ============================================================================
\begin{figure}\caption[]
{Borderlines $r_{c-a}$ between absolute and convective instability versus
through-flow rate. Symbols and {\it thick} curves represent $r_{c-a}$ obtained
from the full field equations and from the GLE, respectively. {\it Thin} curves
show stability curves
$r_c$ discussed in Sec.~\ref{iv}. The different types of perturbations are 
identified by circles and dotted lines ($S$), upwards pointing
triangles and full lines ($U$),
and downwards pointing triangles and dashed lines ($D$).
The case of a pure fluid, $\psi=0$,
is shown by dash-dotted lines: thick one for $r_{c-a}^S$ and thin one for $r_c^S$.
Parameters are $L=0.01,~\sigma=10,$ and $\psi$ as shown.}
\label{konab}
\end{figure}
% ============================================================================

%\end   {multicols}
\end{document}